\documentclass[prc,twocolumn,amssymb,aps,prc,longbibliography,superscriptaddress]{revtex4-1}
\usepackage{amssymb}
\usepackage{graphicx}
\usepackage{dcolumn}
\usepackage{amsmath}
\usepackage{amsfonts}
\usepackage{color}
\usepackage{multirow}
\usepackage{bm}
\begin{document}

\title{Invariant-mass spectroscopy of $^{14}$O excited states}
\author{R.J. Charity}
\affiliation{Department of Chemistry,
Washington University, St. Louis, Missouri 63130, USA}
\author{K.W. Brown}
\affiliation{Department of Chemistry,
Washington University, St. Louis, Missouri 63130, USA}
\author{J. Oko{\l}owicz}
\affiliation{Institute of Nuclear Physics, Polish Academy of Sciences, Radzikowskiego 152, PL-31342 Krak{\'o}w, Poland}
\author{ M. P{\l}oszajczak}
\affiliation{Grand Acc\'el\'erateur National d'Ions Lourds (GANIL), CEA/DSM - CNRS/IN2P3,
BP 55027, F-14076 Caen Cedex, France}
\author{J.M. Elson}
\affiliation{Department of Chemistry,
Washington University, St. Louis, Missouri 63130, USA}
\author{W. Reviol}
\affiliation{Department of Chemistry,
Washington University, St. Louis, Missouri 63130, USA}
\author{L.G. Sobotka}
\affiliation{Department of Chemistry,
Washington University, St. Louis, Missouri 63130, USA}
\author{W.W. Buhro}
\affiliation{National Superconducting Cyclotron Laboratory and 
Department of Physics and Astronomy, Michigan State University, East Lansing, MI 48824, USA}
\author{Z. Chajecki}
\affiliation{National Superconducting Cyclotron Laboratory and 
Department of Physics and Astronomy, Michigan State University, East Lansing, MI 48824, USA}
\author{W.G. Lynch}
\affiliation{National Superconducting Cyclotron Laboratory and 
Department of Physics and Astronomy, Michigan State University, East Lansing, MI 48824, USA}
\author{J. Manfredi}
\affiliation{National Superconducting Cyclotron Laboratory and 
Department of Physics and Astronomy, Michigan State University, East Lansing, MI 48824, USA}
\author{R. Shane}
\affiliation{National Superconducting Cyclotron Laboratory and 
Department of Physics and Astronomy, Michigan State University, East Lansing, MI 48824, USA}
\author{R.H. Showalter}
\affiliation{National Superconducting Cyclotron Laboratory and 
Department of Physics and Astronomy, Michigan State University, East Lansing, MI 48824, USA}
\author{M.B. Tsang}
\affiliation{National Superconducting Cyclotron Laboratory and 
Department of Physics and Astronomy, Michigan State University, East Lansing, MI 48824, USA}
\author{D. Weisshaar}
\affiliation{National Superconducting Cyclotron Laboratory and 
Department of Physics and Astronomy, Michigan State University, East Lansing, MI 48824, USA}
\author{J.R. Winkelbauer}
\affiliation{National Superconducting Cyclotron Laboratory and 
Department of Physics and Astronomy, Michigan State University, East Lansing, MI 48824, USA}
\author{S. Bedoor}
 \altaffiliation[Present address: ]{Physics Department, Kettering University, Flint, MI 48504, USA}
\affiliation{Department of Physics,
Western Michigan University, Kalamazoo, Michigan 49008, USA}

\author{A.H. Wuosmaa}
\affiliation{Department of Physics,
Western Michigan University, Kalamazoo, Michigan 49008, USA}
\affiliation{Department of Physics, University of Connecticut, Storrs, Connecticut 06269, USA}

\date{\today}

\begin{abstract}
Excited states in $^{14}$O have been investigated both experimentally and theoretically. Experimentally, these states were produced via neutron-knockout reactions with a fast $^{15}$O beam  and the invariant-mass technique was employed  
to isolate the 1$p$ and 2$p$ decay channels and determine their branching ratios.
The spectrum of excited states was also  calculated with the Shell Model Embedded in the Continuum that treats bound and scattering states in a unified model.  
By comparing energies, widths and decay branching patterns, spin and parity assignments for all experimentally observed levels below 8~MeV are made. This includes the location of the second 2$^{+}$ state that we find is in near degeneracy with the third 0$^{+}$ state.  An interesting case of sequential 2$p$ decay through a pair of  degenerate $^{13}$N excited states with opposite parities was found where the interference between the two sequential decay pathways produces an unusual relative-angle distribution between the emitted protons.

\end{abstract}

\maketitle
\section{INTRODUCTION}
The past decade has been one in which nuclear structure theory has seriously attacked the treatment of the effects of open channels.   Clearly this advance is required to splice structure and reaction theories.  Thus far, the lessons from this effort are many and include an explanation of several levels close to decay thresholds  \cite{smec2,smec3,oko2016,oko2018} and the demonstration of the enhanced mixing of standard shell-model eigenstates when the coupling to the continua is allowed \cite{smec,smec2,smec3}. Both of these features can destroy touchstones, like mirror symmetry, employed by nuclear physicists for decades. 

The present work contains a combined experimental - theoretical study of  $^{14}$O.  All excited states of this nucleus can decay by 1$p$ emission and 2$p$ emission becomes possible above 6.6 MeV of excitation for which several excited states of  $^{13}$N are possible intermediates. The branching between these possible decay routes is accessible by invariant-mass spectroscopy, the experimental tool employed here. Comparing energies, widths and branching information to Shell Model Embedded in the Continuum (SMEC) calculations provide evidence that the second 2$^{+}$ state has an almost degenerate partner for which only the third 0$^{+}$ state is fully consistent with the SMEC calculations. Having made this assignment, there is little ambiguity as to the assignment of the third 2$^{+}$ state and assignments to other high-lying levels are considered. 

\section{EXPERIMENTAL METHODS}

The experiment was performed at the Coupled Cyclotron Facility at the National Superconducting Cyclotron Laboratory at Michigan State University.
Details have been published elsewhere \cite{Brown:2014,Brown:2015,Brown:2017}. Briefly, a mixed $^{17}$Ne ($E/A$=62.9~MeV, 11\%, 1.6$\times$10$^{4}$ pps) and $^{15}$O($E/A$=52.1~MeV, 89\%) secondary beam was produced from a 
primary  $^{20}$Ne beam ($E/A$=150 MeV, 175 pnA). This secondary beam impinged 
 on a 1-mm-thick $^9$Be target and charged particles produced in the subsequent reactions were detected in the High Resolution Array (HiRA) \cite{Wallace:2007}. In this experiment, the array  consisted of  fourteen $E$-$\Delta E$  telescopes which were arranged around the beam axis
and subtended polar angles from 2$^\circ$ to 13.9$^\circ$. $^{13}$N and $^{12}$C fragments were only identified in the inner two telescopes where their yield was greatest.

The CsI(Tl) $E$ detectors in each telescope were calibrated using $E/A$=55 and 75-MeV proton and $E/A$=55 and 75-MeV  $N$=$Z$ cocktail beams. The $^{13}$N calibration was constructed from the $^{14}$N calibrations. The relative locations of the target and each HiRA telescope were determined accurately using a coordinate measurement machine arm.  Using the same experimental data, we have examined narrow one- and two-proton resonances in $^{12-15}$N, $^{13}$O and $^{15}$O whose centroids are well constrained. The invariant-mass peaks were found to be less than 10~keV from their tabulated values \cite{Solove:1991} and so we use 10~keV as the systematic uncertainty in extracting excitation energies.

The $\gamma$-ray array CAESAR \cite{Weisshaar:2010} surrounded the target to detect any $\gamma$ rays in coincidence with the detected charged particles. For this experiment, the array consisted of 158 CsI(Na) crystals covering polar angles between 57.5$^\circ$ and 142.4$^{\circ}$ in the laboratory, with complete azimuthal coverage. The only use of this array in this study was to verify that the invariant-mass peaks obtained with the detected 2$p$+$^{12}$C channel involved decays to the ground state of $^{12}$C rather than the $\gamma$-decaying first-excited state.

For the normalization of cross sections, the number of beam particles was determined by counting using a thin plastic-scintillator foil placed in the focal point of the A1900 separator. The loss in the beam flux in its transportation to the target and the relative contribution of each beam species was determined by temporarily placing a CsI(Tl) detector in the target position. As we are only interested in relative yields, the uncertainties in the cross sections quoted in this work are statistical only. However, based on past experience \cite{Charity:2019} we expect an overall $\pm$15\% systematic uncertainty.  

\section{Simulations}
\label{sec:sim}
In order to extract intrinsic widths and cross sections, it is important to understand the experimental invariant-mass resolution and the detection efficiency. To explore this, we have performed Monte Carlo simulations including the detector geometry and their energy and position resolutions. The major contribution to the energy resolution comes from the energy loss and small-angle scattering of the decay fragments as they leave the target which are calculated in the simulations using Refs.~\cite{Ziegler:1985,Anne:1988}. The reaction is assumed to occur at a random depth within the $^9$Be target.  The primary angular distribution of $^{14}$O$^*$ fragments produced in the reactions is chosen so as to give ``detected'' secondary distributions in the simulations that are consistent with the  experimental distributions. 

\begin{figure}[tbp]
\includegraphics[scale=.43]{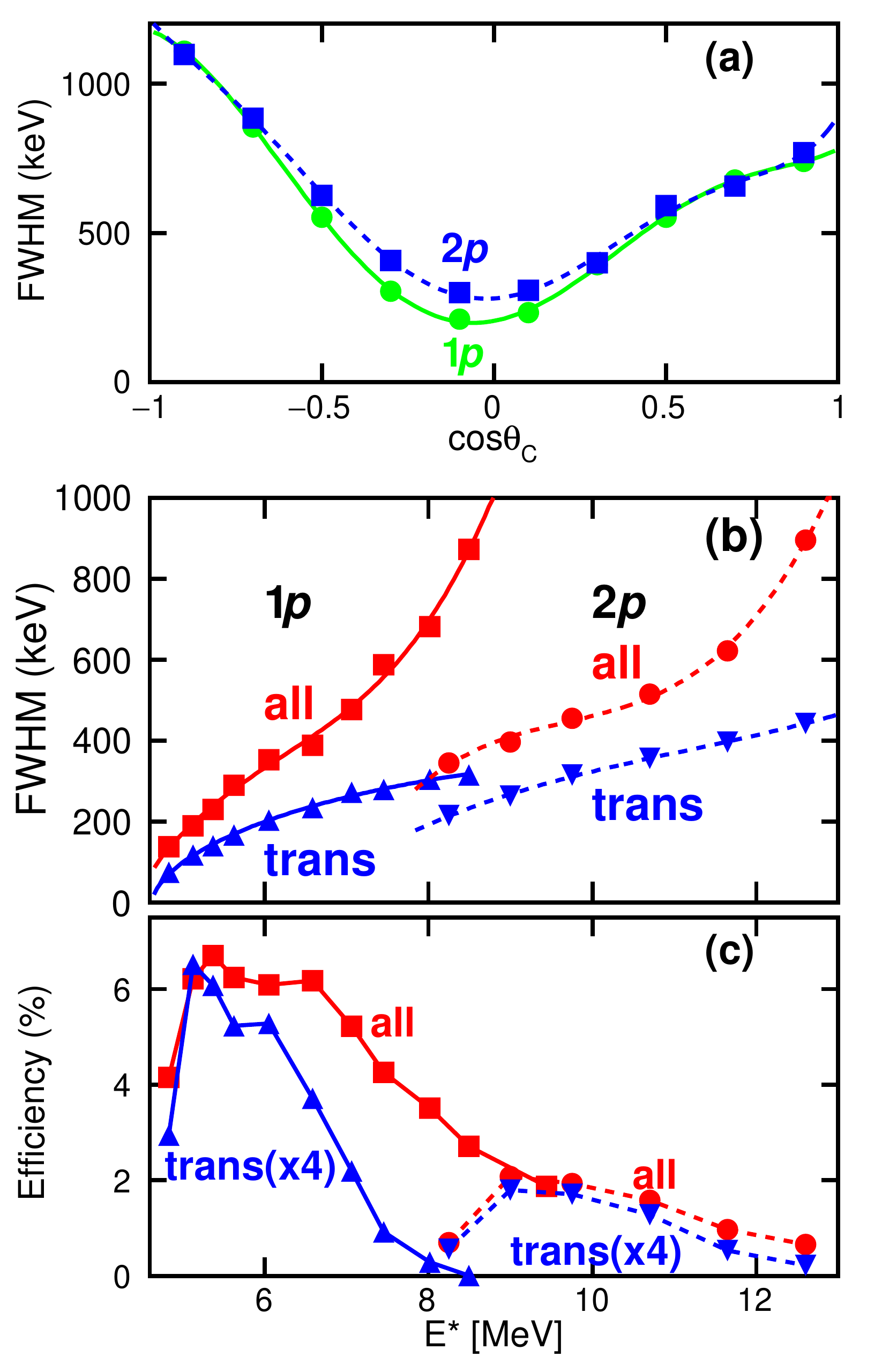}
\caption{Simulated invariant-mass resolution for 1$p$ and 2$p$ emissions from $^{14}$O. Panel(a) shows the dependence of the resolution on the emission angle $\theta_C$ of the heavy core in the $^{14}$O$^*$ center-of-mass frame. Dependence of the (b) resolution and (c) detection efficiency on the $^{14}$O$^*$ excitation energy for all and only transverse emissions of the core. Note that the efficiency for transverse emissions has been multiplied by a factor of 4 for display purposes.}
\label{fig:res}
\end{figure}

Figure~\ref{fig:res}(a) shows the predicted energy resolution as a function of the emission angle $\theta_{C}$ of the remaining heavy ``core'' fragment in the $^{14}$O$^*$ center-of-mass frame for a reaction induced by a $^{15}$O beam particle. Here zero degrees is the beam axis. Results are shown for simulations of 1$p$  and 2$p$ decay where the excitation energies are chosen to match the 2$^{+}_{1}$ and 2$^{+}_{3}$ states of $^{14}$O \cite{Solove:1991}, respectively. The protons are assumed to be emitted isotropically in the $^{14}$O$^*$ center-of-mass and the 2$p$ decay is treated as sequential emission though the 5/2$^{+}$, third-excited state of $^{13}$N. The resolution, as measured by the FWHM of the response function, shows a similarly strong dependence on $\theta_{C}$ for both simulations. This behavior is due to the use of a thick target.
 
In analyzing both the experimental and simulated events, the energies of the detected  fragments are increased to account for their average energy loss in the target material. These increases are calculated assuming the reaction occurred in the center of the target, but in reality this could be anywhere between the front and back of the target. For a 1$p$ decay, the differential velocity loss in the target material is $\approx$5 times larger for the $^{13}$N fragment than for the emitted proton \cite{Ziegler:1985}. Thus the uncertainty in the velocity of  the heavy fragment  is most important in determining the resolution.  For transverse decay, {\em i.e.}, where the $p$-$^{13}$N relative velocity vector is perpendicular to the beam axis, this uncertainty in the $^{13}$N laboratory velocity acts perpendicularly to the relative velocity and thus only affects its value  in second order. The final resolution is dominated by  small-angle scattering of the fragments in the target material. However for longitudinal decay, the uncertainty due to the energy loss acts in first order while that from the small-angle scattering acts in second order. The observed angular dependence therefore reflects the relatively larger importance of the energy-loss contribution over the small-angle-scattering contribution to the final resolution. Clearly, improved resolution can be obtained from selecting events associated with transverse emission. From here on we will define these as events with $\left|\cos\theta_{C}\right|<$0.2. 

Figures~\ref{fig:res}(b) and \ref{fig:res}(c) shows the simulated resolution and detection efficiency as a function of excitation energy. Results are shown for all ``detected'' events and those in our transverse gate. The improved resolution from the transverse gate comes at the cost of about a factor of 4 reduction in the efficiency. The resolution is best close to the threshold and deteriorates  with increasing excitation energy.
The rapid increase in the FWHM at the higher excitation energies for events without any gate, is due to the loss of transverse events, {\em i.e.} the efficiency for transverse emission drops as the typical opening angles of the fragments in the laboratory frame become larger than the angular acceptance of the detector. 
 
Apart from the improved resolution, there are two other reasons to use the transverse events to fit the excitation energies and widths.
First, the determination of the experimental resolution is more certain for the transverse events. In determining an average  resolution for all detected events, one needs to know the relative number of events as a function $\theta_{C}$. This  dependents on the intrinsic $\theta_{C}$ distribution and  the relative detection efficiency as a function of $\theta_{C}$. The latter depends of the assumed primary-fragment angular distribution and the placement of the detectors. By using only a small range of $\theta_{C}$ values these uncertainties are eliminated. Second, the effect of errors in the CsI(Tl) calibration for 
the heavy fragments are minimized. For transverse emission such errors, like the energy-loss correction, act perpendicular to the relative velocity and thus only contribute in 2\textsuperscript{nd} order.

\section{EXPERIMENTAL RESULTS}
\begin{figure*}[tbp]
\includegraphics[scale=.92]{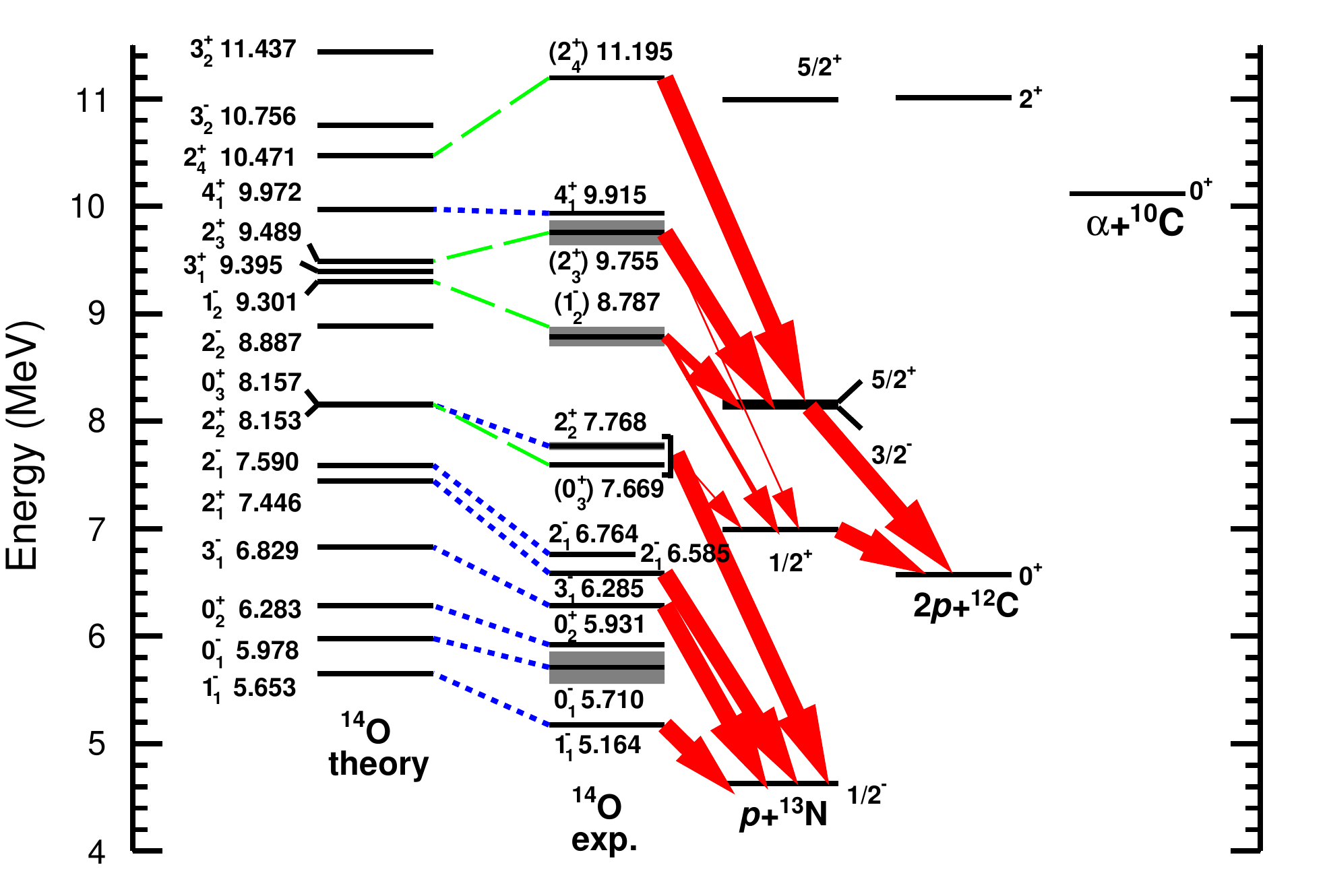}
\caption{Comparison of the experimental and SMEC level schemes for $^{14}$O.
The experimental level scheme is derived from the most recent evaluation \cite{Solove:1991} and from the states observed in this work and Refs.~\cite{Teranishi:2007,Wang:2008}. Observed proton decay branches are indicated by the arrows. The width of the arrows is roughly proportional to relative intensity of the decay branch. The excitation energies of the  levels are given in MeV. Note it is possible that some of the higher-lying states have significant unobserved decay branches to $^{13}$N$_{g.s.}$ (see Table~\ref{tab:level}). The SMEC results are obtained using the modified YSOX interaction \cite{yuan2012} and the Wigner-Bartlett continuum-coupling interaction with the continuum-coupling strength $V_0 = -350$ MeV$\cdot$fm$^3$. For details of the calculation, see Sec. \ref{sec:theory}. The experimental levels whose spin and party were assigned before this work are connected by dotted blue lines to their theoretical partners while the assignments make or confirmed in this work are connected by the dashed green lines. }
\label{fig:level}
\end{figure*}

\begin{figure}[tbp]
\includegraphics[scale=.43]{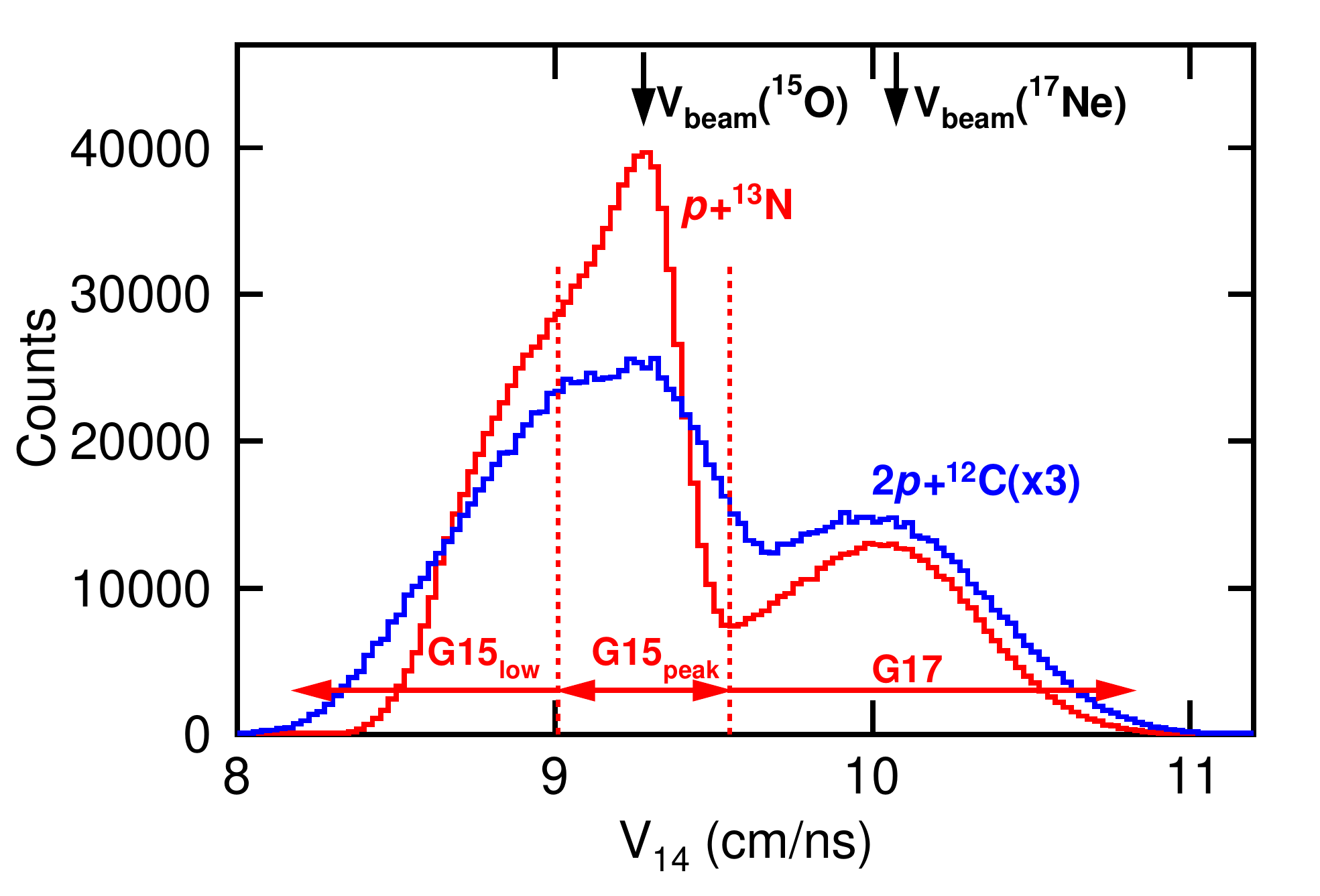}
\caption{Laboratory velocity distribution of the parent $^{14}$O$^*$ fragments reconstructed from the detected $p$+$^{13}$N and 2$p$+$^{12}$C events. Values of the velocities of the $^{15}$O and $^{17}$Ne beam particles are indicated for comparison. Gates $G17$, $G15_{peak}$, and $G15_{low}$ used to examine the $p$+$^{13}$N events are indicated. }
\label{fig:vel}
\end{figure}

Values of the excitation energy, widths, and cross sections for $^{14}$O levels determined in this work are listed in Table~\ref{tab:level} and Fig.~\ref{fig:level} displays the energy level diagram and observed decay branches.  Excited $^{14}$O fragments were produced from reactions induced with both the $^{15}$O and $^{17}$Ne beam particles.  This is illustrated from the distribution of the center-of-mass velocity of the  detected  $p$+$^{13}$N and 2$p$+$^{12}$C events which is shown in Fig.~\ref{fig:vel}. Both distributions contain two peaks, with their maxima close to the velocities of the two beam species. Note that the low-energy particle-identification thresholds of the 
CsI(Tl) detectors severely restricts events below $\approx$8.5 cm/ns. 
We have also looked for peaks in the $\alpha$+$^{10}$C invariant-mass distribution as a number of clustered states are predicted just above the threshold for this channel \cite{Baba:2019}. However, no resolved states were observed.

\begin{table*}
\caption{States in $^{14}$O investigated in this work and their extracted properties. For the fitted excitation energies, only the statistical errors are listed, the systematic uncertainty is estimated to be $\pm$10~keV. For comparison, excitation energies ($E^*_{eval.}$) from the most recent evaluation \cite{Solove:1991} are also listed. Cross sections are listed for decays to the ground-state ($p_{0}$),  first-excited state ($p_{1}$), and the 5/2$^{+}$,3/2$^{-}$ doublet ($p_{2}+p_{3}$) of states in $^{13}$N.The theory-based  spin-parity assignments made or confirmed in this work are shown in bold font.}
\begin{ruledtabular}
\begin{tabular} {ccccccc}
$E^*$ & $\Gamma$  & $J^{\pi}$\footnotemark[3] & $\sigma$($p_0$) & $\sigma$($p_1)$ & $\sigma$($p_2+p_3$) & $E^*_{eval.}$\\
{[}MeV] & [keV] &  & [mb] & [mb] & [mb] & [MeV]\\
\hline
 5.164(2) &  & 1$^-_1$ & 0.25(2) & & & 5.173(10)\\

 6.285(2) &  &3$^-_1$ & 0.37(14) & & & 6.272(10)\\

 6.585(1) & $<$25 &  2$^{+}_{1}$ & 3.46(5) & & &  6.590(10)\\

7.669(53) & $<$128       & $\mathbf{(0^+_3)}$   &             \multirow{2}{*}{\bigg\}3.2(2)} & \multirow{2}{*}{\bigg\}0.083(12)\footnotemark[2]}\\

 7.768\footnotemark[1] & 76\footnotemark[1] & 2$^{+}_{2}$ & & & & 7.768(10)\\

 8.787(13) & 182(32) &  $\mathbf{(1^-_2)}$ & $<$1.0 & 0.16(5) & 0.37(12)& 8.720(40)\\

 9.755(10) & 229(51)  & $\mathbf{(2^{+}_{3})}$ & $<$0.7 & 0.21(2) & 3.11(36) & 9.715(20)\\

11.195(30) & $<$220  &$\mathbf{(2^+_4)}$  & $<$3.4 &   $<$0.25         &      0.78(35)         & 11.240(50)           \\
\end{tabular}
\end{ruledtabular}
\footnotetext[1]{Fixed to value from \cite{Solove:1991}.}
\footnotetext[2]{For $E^*<$8.3~MeV.}
\footnotetext[3]{Bold characters for spin-parities indicate that the assignment is new.}
\label{tab:level}
\end{table*}

\subsection{$p$+$^{13}$N Events}
\label{sec:p13N}

\subsubsection{Excitation-energy distributions}

\begin{figure}[tbp]
\includegraphics[scale=.43]{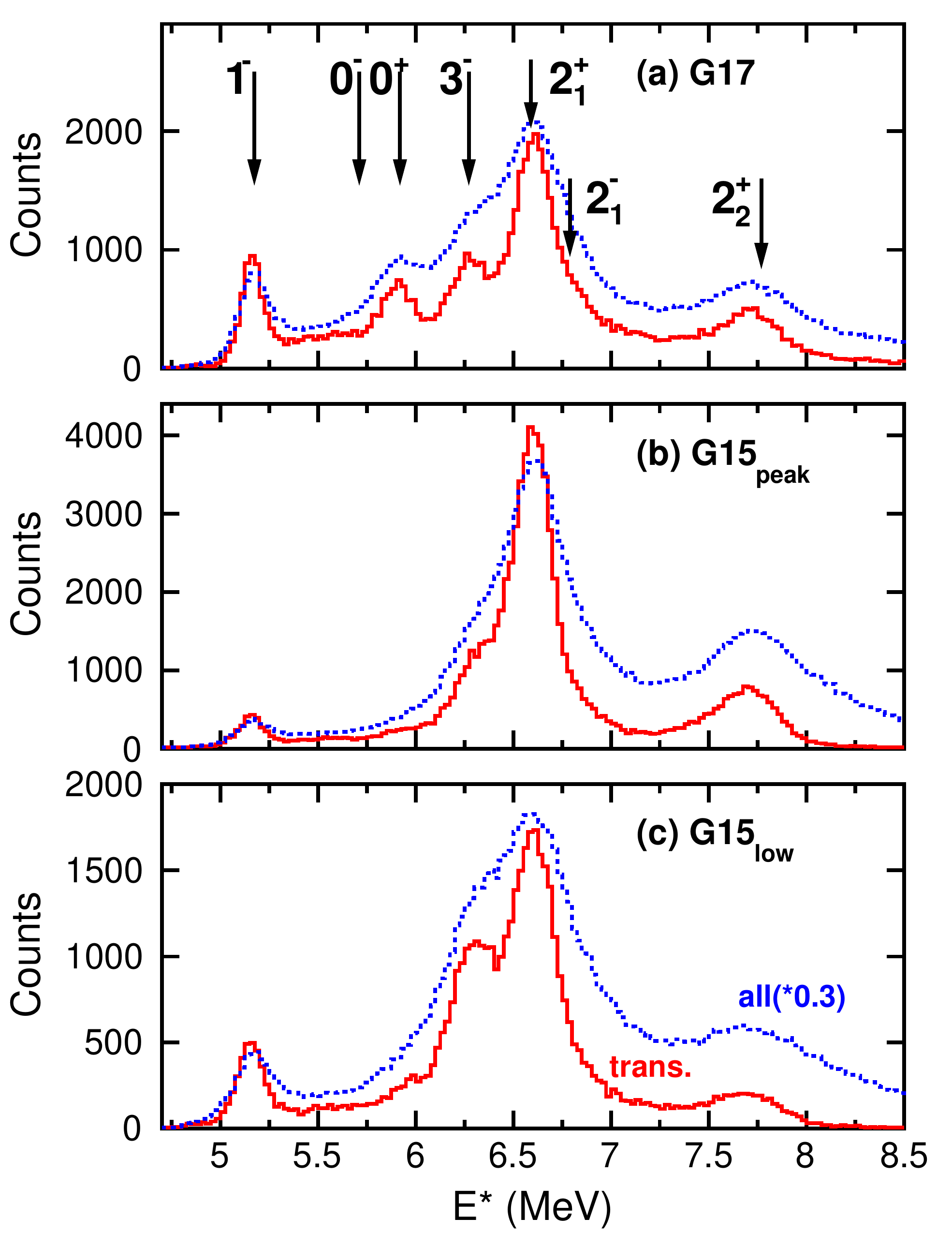}
\caption{$^{14}$O excitation-energy distributions from detected $p$+$^{13}$N events. Solid red histograms are where the $^{13}$N fragment is emitted transversely from the parent $^{14}$O$^*$ system while the dotted blue histograms are for all events scaled by the factor 0.3. (a) gated on the $^{17}$Ne beam (gate $G17$), (b,c) on the $^{15}$O beam. (b) is for $^{14}$O velocity gate $G15_{peak}$ around the peak, while (c) is for the $G15_{low}$ gate containing the low-velocity tail. The arrows in (a) show the locations of states listed in the most recent evaluation \cite{Solove:1991} and from \cite{Teranishi:2007,Wang:2008}.}  
\label{fig:p13N}
\end{figure}

Excitation-energy distributions for the $p$+$^{13}$N channel determined with the invariant-mass technique are shown in Fig.~\ref{fig:p13N} for three gates on the parent $^{14}$O$^*$ velocity. As there are no particle-bound excited states in $^{13}$N, all
single-proton decays are to the ground-state of $^{13}$N.
The improved resolution for transverse events (solid red histograms) is easily seen compared to those for all events (dotted blue histograms).
The reconstructed $^{14}$O excitation-energy distribution for events induced with the $^{17}$Ne beam (gate $G17$) is displayed in Fig.~\ref{fig:p13N}(a). The transverse-emission spectrum shows the largest number of resolved peaks and these can be identified by the arrows in this figure at the energies listed in the most recent evaluation \cite{Solove:1991} and for the $J$=0$^-_1$ and 2$^{-}_{1}$ states from \cite{Teranishi:2007,Wang:2008}.  The mechanism for producing $^{14}$O states from $^{17}$Ne probably involves the knockout of one to three nucleons producing states which sequential decay to $^{14}$O$^*$.

We expect single-neutron knockout to be the most important mechanism with the  $^{15}$O beam especially for $^{14}$O$^*$ velocities near to the $^{15}$O  beam velocity. The neutron configuration of $^{15}$O is well described by a hole in the $p$ shell and thus $p$-shell knockout will produce $J^{\pi}$=0$^+$, 1$^+$, and 2$^+$ states. The excitation-energy spectrum gated around the $^{15}$O beam velocity (gate $G15_{peak}$ in Fig.~\ref{fig:vel}) shown in Fig.~\ref{fig:p13N}(b) is dominated by a peak at the first 2$^+$ energy and a contribution from a peak near the energy of the second 2$^+$ state is also visible. The lowest-energy peak is the $1^-_{1}$ state which could be produced from knockout of a deeply bound  0$s_{1/2}$ neutron. Also present is a low-energy shoulder on the 2$^+_1$ peak which can be attributed to the 3$^-_1$ state. This cannot be the result of simple neutron knockout reaction, and suggests more complicated multi-step processes are contributing.  Such multi-step processes are expected to be more dominant at lower $^{14}$O velocities as verified in Fig.~\ref{fig:p13N}(c) which is gated on the low-velocity tail of the $^{15}$O-induced  events , {\em i.e.} gate $G15_{low}$ in Fig.~\ref{fig:vel}. Here we see the relative contribution of the 1$^-_1$ and 3$^-_1$ peaks are enhanced significantly.

\begin{figure}[tbp]
\includegraphics[scale=.43]{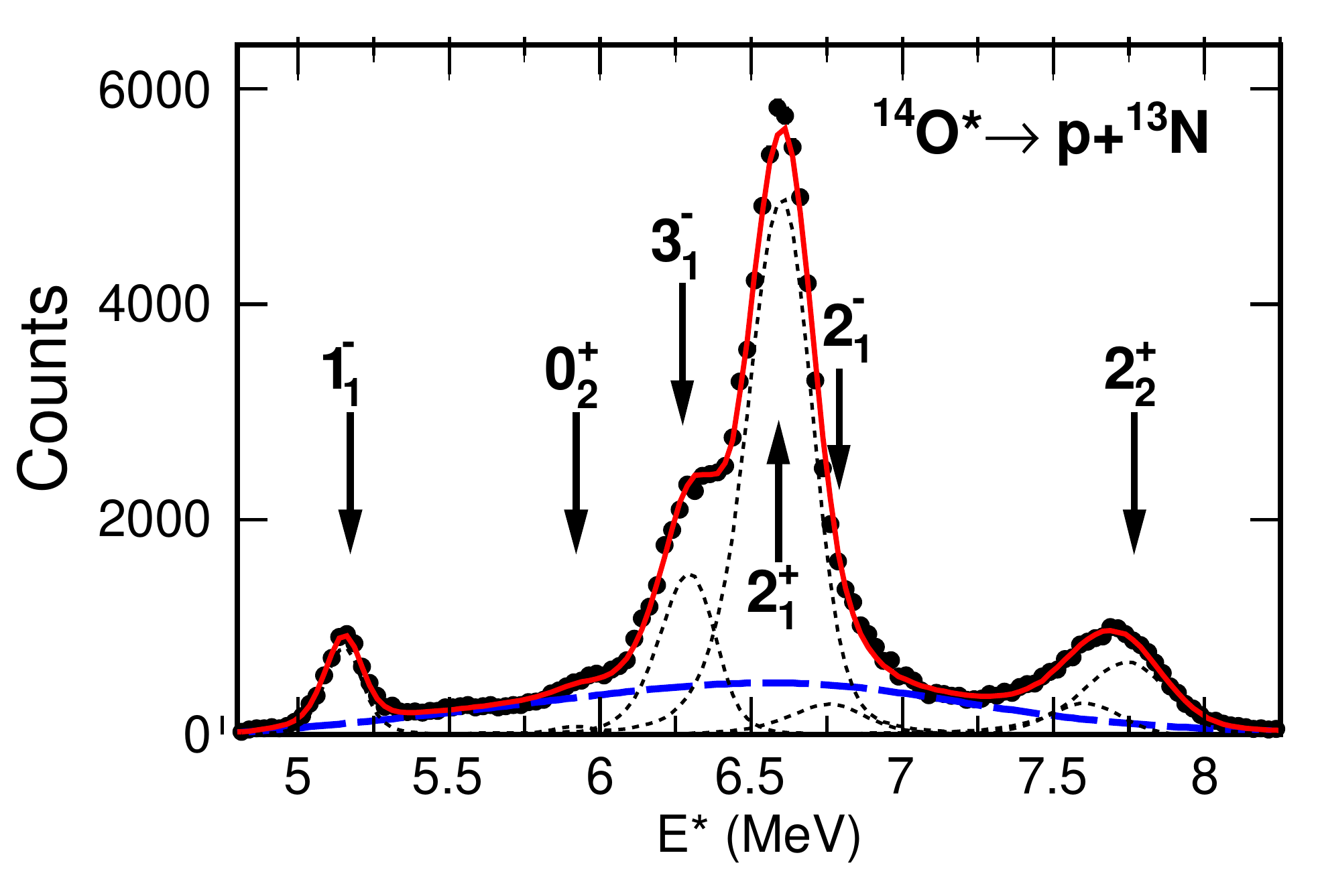}
\caption{Fit to the excitation-energy distribution for transverse $p$+$^{13}$N events with the combined $G15_{peak}$ and $G15_{low}$ gates. Individual curves (dotted) are shown for each state included in the fit and the fitted background distribution is shown as the dashed blue  curve. Arrows indicate the locations of states listed in the most recent evaluation \cite{Solove:1991} and from \cite{Teranishi:2007,Wang:2008}. } 
\label{fig:fitp13N}
\end{figure}

We will concentrate on fitting the centroids, widths and cross sections determined with the $^{15}$O beam as the statistics are better.  For the yields, we have fit the combined $G15_{low}$ and $G15_{peak}$ gates. However, in fitting of the centroids and widths of the 2$^+_1$ and 2$^{+}_2$ peaks, we have just used the $G15_{peak}$ gate as contamination from the other peaks, such as the 3$^-_1$, is minimal. 
Figure~\ref{fig:fitp13N} shows a fit to this combined $G15_{low}+G15_{peak}$ spectrum using Breit-Wigner line shapes and a smooth background (dashed curve).
Backgrounds are typical needed in fitting invariant-mass spectra and, in general, represent contributions from non-resonant breakup and unresolved wide states. However in this case, we do not expect any unresolved wide states in this interval of excitation energy. 
  Our Monte Carlo simulations were used to incorporate the experimental resolution and the fitted contribution from individual levels are shown by the dotted curves. This fit includes  peaks for the 1$^-_1$, 0$^+_2$, 3$^{-}_1$, 2$^{+}_{1}$, 2$^{-}_{1}$, and 2$^{+}_2$ states. Other than the highest-energy peak which will be discussed later, only the 1$^-_1$, 3$^-_1$ and 2$^+_1$ peaks have  significant yields in the fits. The fitted centroids of these levels in Table~\ref{tab:level} are consistent with the evaluated values  \cite{Solove:1991} within our statistical and systematic uncertainties. 

The intrinsic width of the 2$^{+}_{1}$ state is of interest as this state sits very close to the 2$p$ threshold.  In such cases, a collective state can be formed which carries many features of the nearby particle-emission threshold  \cite{smec2,smec3}. A well know case is  $^{8}$Be$_{g.s.}$ which has strong 2$\alpha$ correlation in its wavefunction due to the presence of the nearby 2$\alpha$ decay threshold. In the present case, the wavefunction should have strong 2$p$ character thus giving it a  small decay width for the  1$p$ channel. We found the fitted width be consistent with zero with an upper limit of 25~keV at the 3$\sigma$ level. This is consistent with the more restrictive limit of $<$5~keV from \cite{Negret:2005}.

\begin{figure}[tbp]
\includegraphics[scale=.43]{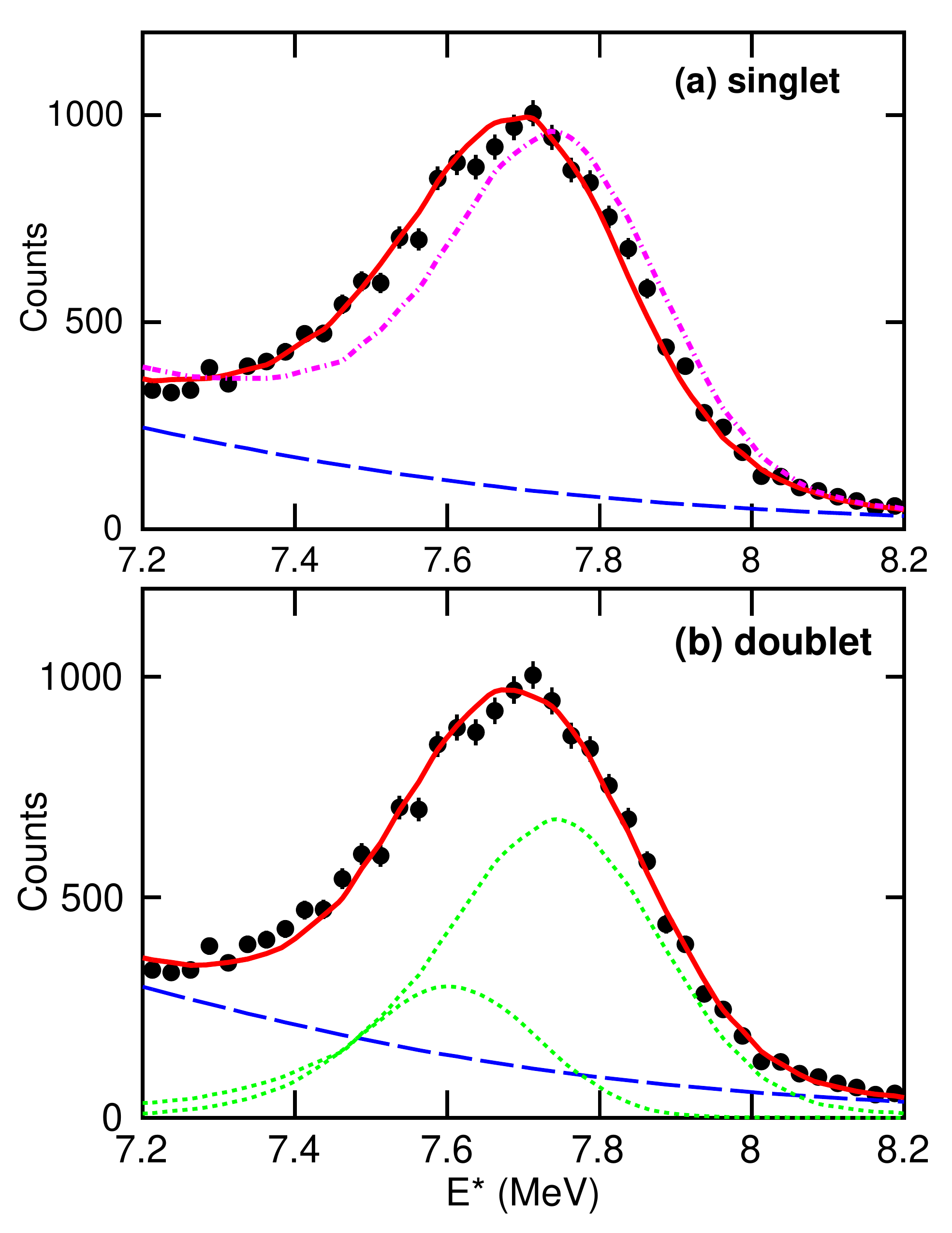}
\caption{Fits to the peak near the location of the known 2$^+_2$ state. The solid red curve in (a) is a fit with a single Breit-Wigner intrinsic line shape where the centroid and width are allowed to vary. The dot-dashed magenta curve was obtained when the centroid and width are constrained to the known values for the 2$^+_2$ state. (b) shows a fit assuming the observed peak is a doublet. The individual peaks are indicated by the dashed lines. The higher-energy peak has its centroid and width constrained to the values for the 2$^+_2$ state. The smooth dashed blue curve in both panels is the fitted background.}
\label{fig:2plus2}
\end{figure}

For the peak at the energy of the 2$^+_2$ state, a good fit  was obtained with a  single peak with a centroid of $E^*$=7.723(2)~MeV and an intrinsic  width of $\Gamma$=128(18)~keV (solid red curve in Fig.~\ref{fig:2plus2}(a)]. However, the latter is almost a factor of two larger than the value of $\Gamma$=76(10)~keV listed in the most recent evaluation \cite{Solove:1991}. More recent measurements give similarly small values; 63(16)~keV \cite{Teranishi:2007} and 62(10)~keV \cite{Wang:2008}.  We therefore conclude that this peak is either not the 2$^+_2$ state or more likely a doublet of the $2^+_2$ and a previously unknown $^{14}$O state.  To further highlight this point, Fig.~\ref{fig:2plus2}(a) also shows a calculation (dot-dashed magenta curve) with the centroid fixed to the 2$^{+}_{2}$ evaluated value and the width fixed to the weighted average of the previously reported values. This calculated peak is shifted to higher excitation energies than the experimental peak and does not reproduce the data very well.

Figure~\ref{fig:2plus2}(b) shows a fit to this peak with a doublet of states. The higher-energy member of the doublet has its centroid and width  constrained to the   2$^+_2$ values [same as for the dot-dashed curve in Fig.~\ref{fig:2plus2}(a)] while the width of the low-energy member is assumed to be zero. As long as the width of this new state is about the same as the 2$^+_2$ level or smaller, rather similar fits are produced. The maximum width of this new state is $\Gamma$=128~keV, i.e. the value obtained from fitting this peak as a singlet.

\subsubsection{Parent velocity distribution}
\label{sec:pCVcm}
\begin{figure}[tbp]
\includegraphics[scale=.43]{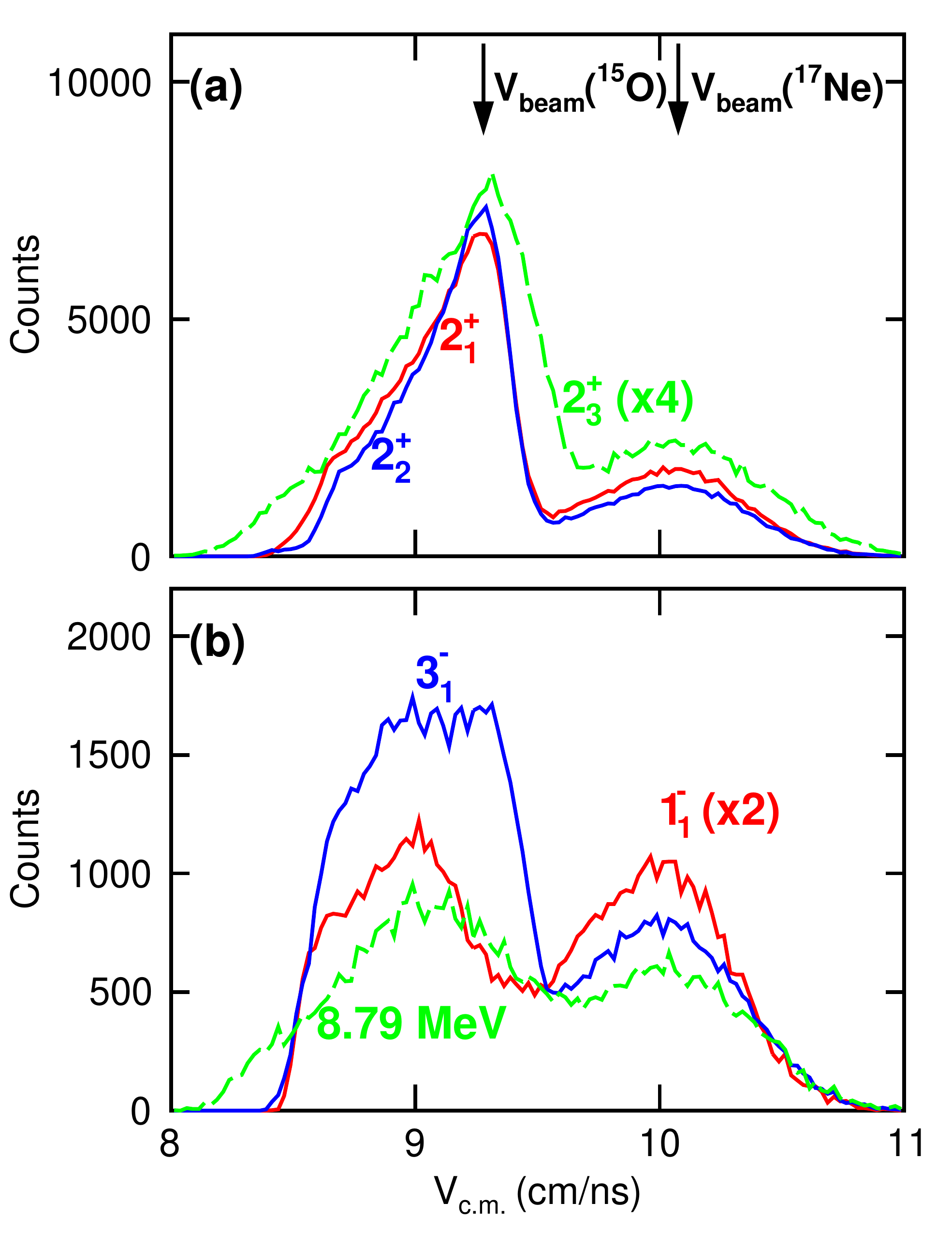}
\caption{Reconstructed $^{14}$O$^*$ parent velocity distributions in the laboratory frame  determined for the indicated peaks. For reference, the two beam velocities are shown by the arrows.}
\label{fig:Vcm}
\end{figure}

To explore the production mechanism for the observed states, Fig.~\ref{fig:Vcm} displays reconstructed $^{14}$O$^*$ velocity spectra gated on some of the different peaks in the excitation spectra. For the 2$^+_1$ peak, we have employed a narrow gate to preclude the possibility of any significant  contribution from the neighboring 3$^-_1$ peak. Using the simulated resolution, we then use this yield to subtract the relevant contribution from this state under the 3$^-_1$ peak. Velocity distributions for  the peaks at the  2$^+_1$ and 2$^+_2$ energies are shown as the solid curves in the Fig.~\ref{fig:Vcm}(a)  while distribution for the 1$^-_1$ and 3$^-_1$ peaks are shown as the solid curves  in Fig.~\ref{fig:Vcm}(b),
The former show very sharp peaks at the $^{15}$O beam velocity expected for single-neutron knockout from a valance state. No dependence of the velocity distribution for the peak at the 2$^+_2$ energy was observed as we scanned the energy across the peak. If this peak is a doublet, then the unknown member also appears to be produced by the knockout of a valence $p$-shell neutron and thus is $J^{\pi}$=0$^+$, 1$^+$, or 2$^{+}$. Indeed the  0$^{+}_{3}$ state is a candidate for this state based on theoretical consideration (see later). The 1$^-_1$ and 3$^-_1$ distributions have broader peaks in this region and the 3$^-_1$ distribution has a flat top, while the 1$^-_1$ is peaked below the $^{15}$O beam velocity. The knockout of a deeply bound $s_{1/2}$ neutron should produce a broader peak which could explain the result for the 1$^-_1$ peak, but as mentioned before, the population mechanism for the 3$^-_1$ peak is not clear.

\subsection{2$p$+$^{12}$C Events}
\label{sec:pp12C}
\subsubsection{Excitation-energy distribution}
\begin{figure}[tbp]
\includegraphics[scale=.43]{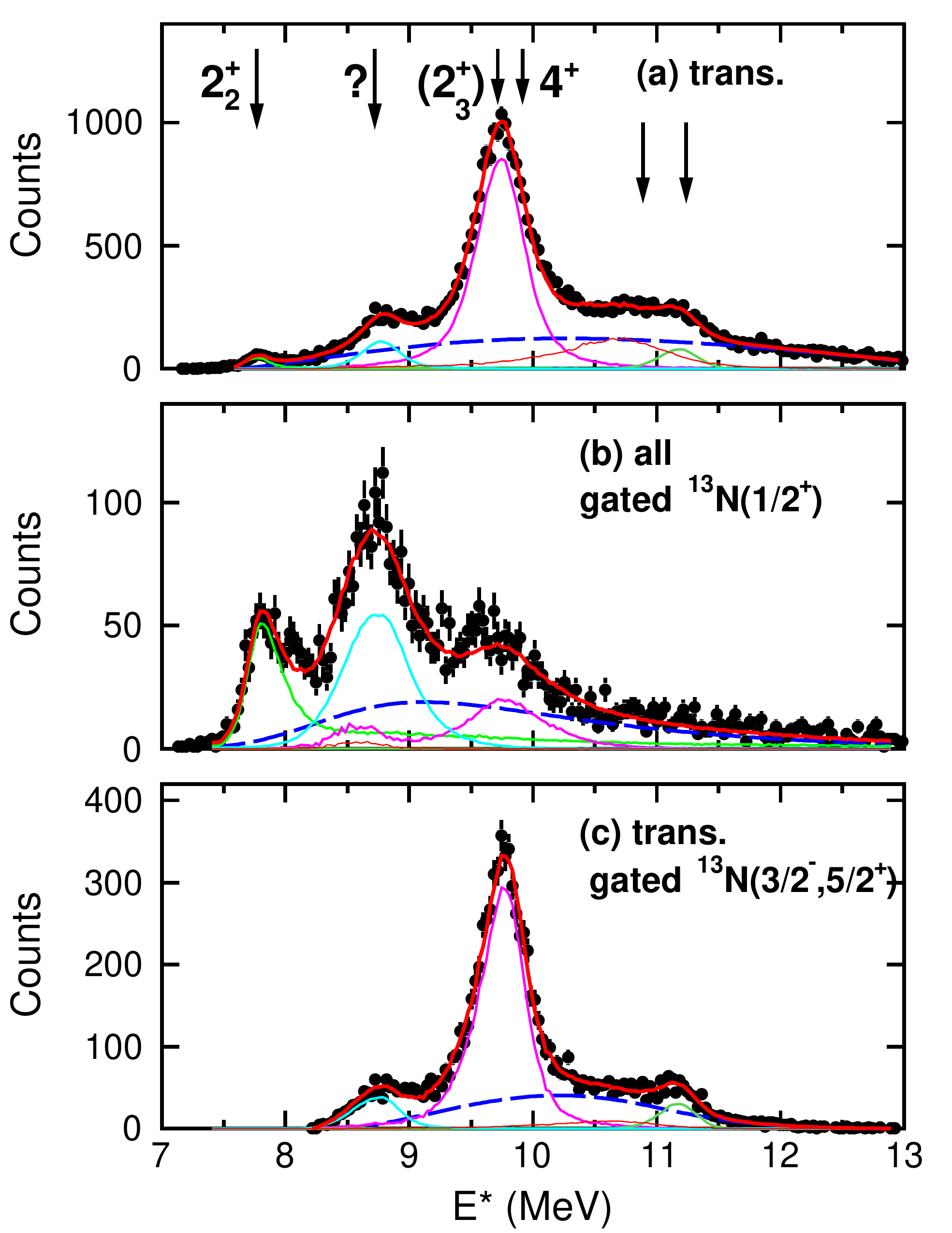}
\caption{$^{14}$O excitation-energy distributions from detected 2$p$+$^{12}$C events. (a) Distribution for all events where the $^{12}$C fragment is emitted transversely. (b) Distribution of all events for the gate on the $J^{\pi}$=1/2$^-$ intermediate state of $^{13}$N. (c) Distribution of transversely-emitted events for the gate on the $J^{\pi}$=3/2$^-$ and 5/2$^+$ intermediate states of $^{13}$N. Curves shows fits to these distribution with the fitted background given by the dashed blue curves and the contribution from the individual states is also shown. The arrows in (a) show the energies  of $^{14}$O levels in the most recent evaluation \cite{Solove:1991}. The labels associated with these indicate the spin and parity assigned in this evaluation when known. The level associated with the question mark was not considered firmly established in the evaluation.}
\label{fig:pp12C}
\end{figure}

The reconstructed excitation-energy distribution for the 2$p$+$^{12}$C events where the $^{12}$C fragment is emitted transversely is shown in Fig.~\ref{fig:pp12C}(a). The peaks observed in this spectrum as associated with 2$p$ decay to the ground state of $^{12}$C. While a small 4.44-MeV $\gamma$-ray yield associated with the decay of the first-excited state of $^{12}$C is present in the coincident $\gamma$-ray spectrum, this yield is not correlated with the invariant-mass peaks and is thus associated with background. This is probably from non-resonant breakup reactions.

This invariant-mass spectrum is dominated by a peak at 9.78~MeV which in the most recent evaluation is tentatively assigned as the 2$^+_{3}$ level. We also observe a small peak just above the 2$p$ threshold near the energy of the $J^{\pi}$=2$^+_2$ state.  The tabulations also tentatively assign a state at 8.72~MeV that corresponds to the small peak  [labeled ``?'' in Fig.~\ref{fig:pp12C}(a)] on the low-energy side of the main 2$^+_3$ peak. Clearly we have confirmed the existence of this state.  In addition there is a structure above the 2$^+_3$ peak which probably has contributions from the 10.89 and 11.2-MeV states listed in the tabulations.   No peaks associated with branches of these levels to 1$p$ channel were observed, possibly due to the low efficiency and poor energy resolution at high-excitation-energy for this channel (Sec.~\ref{sec:sim}). However, Table~\ref{tab:level} provides upper limits to their 1$p$ cross section.

\begin{figure}[tbp]
\includegraphics[scale=.22]{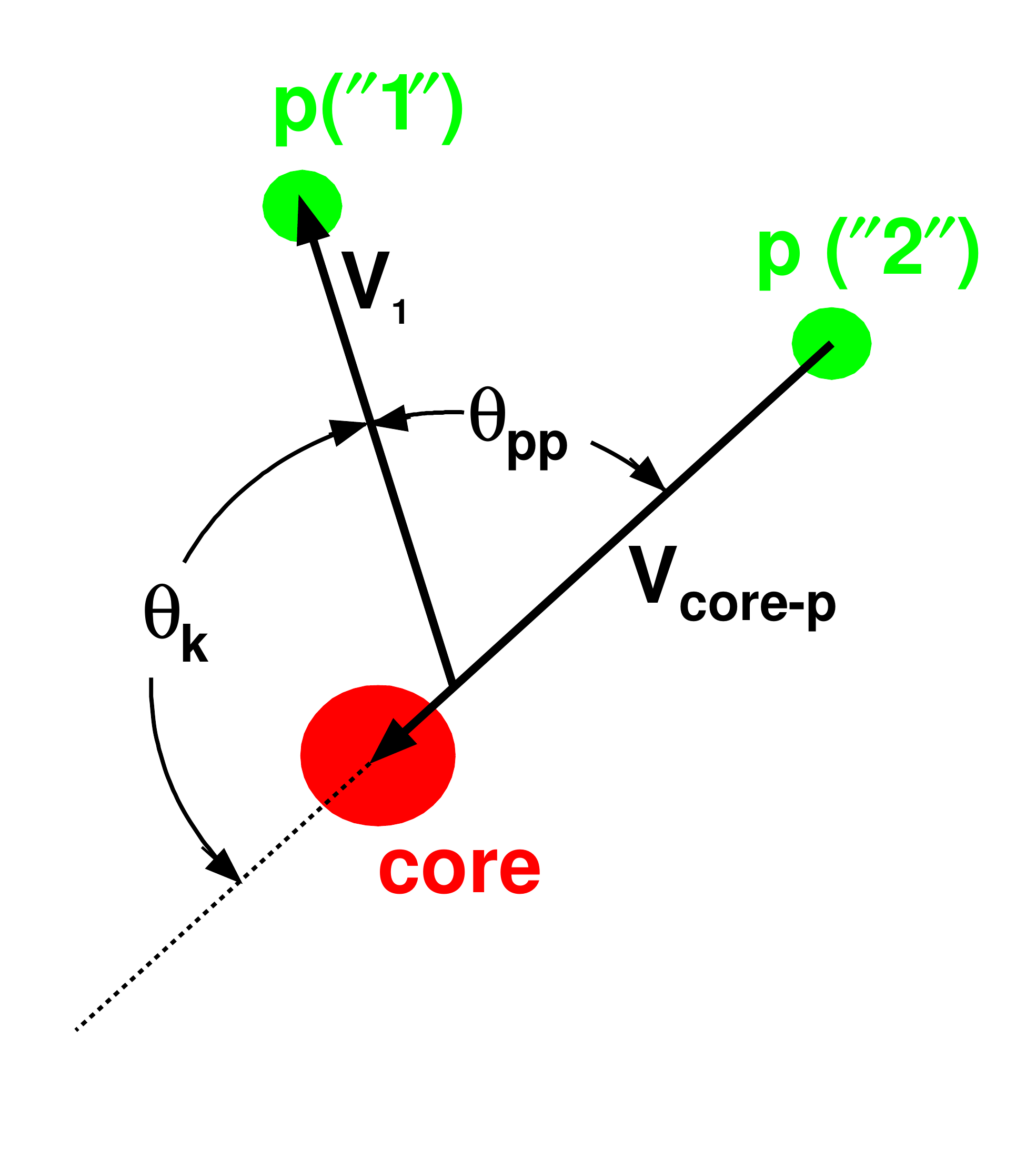}
\caption{Schematic showing the definition of the angle $\theta_k$ used in the Jacobi Y correlation plots. $\bm{V}_{p-core}$ is the relative velocity between proton ``2'' and the core while $\bm{V}_1$ is the relative velocity between proton ``1'' and the center of mass of the other two particles. Also shown is the relationship of $\theta_k$ to the relative angle $\theta_{pp}$. }
\label{fig:jacy}
\end{figure}

\begin{figure}[tbp]
\includegraphics[scale=.43]{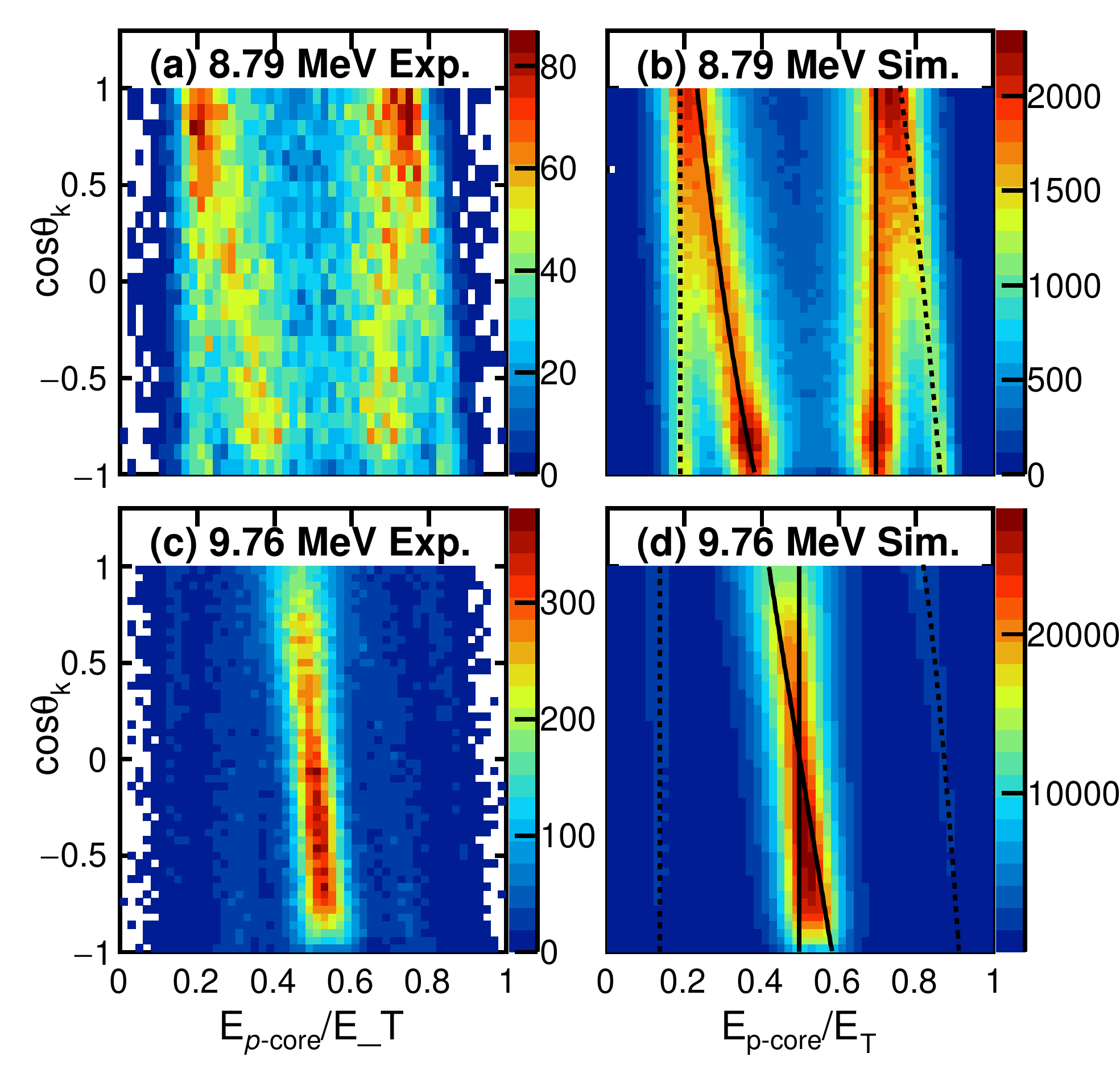}
\caption{Jacobi Y correlation plots for the 2$p$ decay of the 8.79-MeV and 
9.78-MeV peak in the 2$p$+$^{12}$C excitation energy distributions.
(a) and (c) are from experimental data while (b) and (d) are the corresponding  simulated distributions. Curves in (b) and (d) identify the ridge tops associated with different decays. Solid curves are for decay through the 3/2$^-$, 5/2$^+$ doublet, while dotted curves are for decays through the 1/2$^+$ singlet.}
\label{fig:Jac}
\end{figure}

As shown in Fig.~\ref{fig:level}, the sequential 2$p$ decay of these observed states can pass through either the  $J^{\pi}$=1/2$^+$  first-excited state or the 3/2$^-$, 5/2$^+$ doublet.  The latter are separated by 45~keV and, as their intrinsic widths are 62 and 47~keV, respectively, cannot be resolved and hence can be considered degenerate.  Information on the 2$p$ decay mechanism can be gleaned from Jacobi 2-dimensional correlation plots. Consider the schematic velocity vectors in Fig.~\ref{fig:jacy} for two-proton decay where we have designated the protons as ``1'' and ``2''. The  relative velocity vector $\bm{V}_{p-core}$ between the core and proton ``2'' can be used to calculate a relative energy $E_{p-core}$.  If $\bm{V}_{1}$ is the velocity vector between proton ``1'' and the center-of-mass of the other two particles, then $\theta_k$ is the angle between $\bm{V}_1$ and $\bm{V}_{p-core}$.
 In the standard Jacobi Y representation, the $x$ axis is $E_{p-core}/E_{T}$ where $E_T$ is the total decay energy and the  $y$ axis is $\cos\theta_k$.  
 For $\cos\theta_k\approx$-1, the relative angle between the protons is small, while for $\cos\theta_k$=1, they are emitted back-to-back in the $^{14}$O$^*$ center-of-mass frame. For each event, the two-dimensional Jacobi Y histogram is incremented twice, using coordinates calculated by assigning each of the two detected protons to be  proton ``2'' in Fig.~\ref{fig:jacy} and the other to be proton ``1''. Experimental correlation plots are shown for the 8.79 and 9.78-MeV states in Figs.~\ref{fig:Jac}(a) and \ref{fig:Jac}(b), respectively.


A signature of a sequential 2$p$ decay through a single intermediate state is the presence of two separated ridges.
One ridge should be vertical associated with the second emitted proton. Here $E_{p-core}$ is determined from the invariant mass of the intermediate state and thus independent of $\theta_k$.  The other ridge, corresponding to the first emitted proton, should  be tilted, as $E_{p-core}$ in this case depends on the recoil momentum imparted to the core from the second emitted proton and is thus $\theta_k$ dependent. Neither of the experimental Jacobi Y plots in Figs.~\ref{fig:Jac}(a) and \ref{fig:Jac}(c) show two cleanly separated ridges, but they can still be understood as sequential. The  correlation plots in Figs.~\ref{fig:Jac}(b) and \ref{fig:Jac}(d) show corresponding simulated results obtained assuming sequential decays through both the 1/2$^-$ singlet and 3/2$^-$, 5/2$^+$ doublet of states in  $^{13}$N with the relative intensities adjusted to reproduce the experimental correlations.

For the 8.79-MeV state, one can discern a total of four ridges in Fig.~\ref{fig:Jac}(a). To help locate these ridges we have drawn curves along the ridge tops in the simulated distribution of Fig.~\ref{fig:Jac}(b). Dashed-curves marking the  ridges for decay through the 1/2$^+$ $^{13}$N state and the solid curves for decay through the 3/2$^-$, 5/2$^+$ doublet. On the other hand for the 9.78-MeV state, we observe one very intense ridge in Fig.~\ref{fig:Jac}(c). This structure can be traced in the simulations [Fig.~\ref{fig:Jac}(d)] to the two overlapping ridges from the decay through the 3/2$^-$, 5/2$^+$ doublet, {\em i.e.} at this excitation energy the two proton energies are approximately equal. A very close examination of the experimental spectrum in Fig.~\ref{fig:Jac}(a) reveals another pair of faint ridges  associated with decay through the 1/2$^+$ intermediate $^{13}$N state [short dashed curves in Fig.~\ref{fig:Jac}(d)].

It would be useful to gate on  $E_{p-core}$ values for the different intermediate states and project out the $^{14}$O invariant-mass spectra.  However for the 8.79-MeV peak in Fig.~\ref{fig:Jac}(a),  ridges from the two sequential decay paths overlap at $\cos\theta_{k}\approx$1. To avoid this problem we have further restricted the events to $\cos\theta_k<$-0.2. With this extra condition, the excitation energy spectra gated on the 1/2$^-$ singlet and 3/2$^-$, 5/2$^+$ doublet are displayed in Fig.~\ref{fig:pp12C}(b) and \ref{fig:pp12C}(c), respectively. The latter is for transverse emission, but due to smaller number of events, the former has no gate on the emission direction of the core.

These spectra show that the 2$p$ decay path of the 2$^+_2$ level and/or its doublet partner passes only through the 1/2$^-$ singlet (as dictated by energy conservation), while the 8.79 and 9.78-MeV states have contributions from both decay paths. The structures above the 9.78-MeV state appear to be predominately associated with the doublet of intermediate states. To quantify these findings we have fitted these spectra with Breit-Wigner line shapes for each $^{14}$O level considered and  use the Monte Carlo simulations to introduce the experimental resolution and apply exactly the same gates as employed for the experimental data in order to extract reliable relative yields for the two decay branches. There is one exception, the lowest-energy peak which is close to threshold may have an asymmetric line shape for this decay branch. We take a limiting $R$-matrix approximation assuming the total width is dominated by decay to the ground state of $^{13}$N and thus can be described by a Breit-Wigner form $B(E^*)$.  Therefore, the line shape for the small branch to the $^{13}$N excited state is  $P_\ell(E^*-S_p) B(E^*)$ where the first term is the $R$-matrix barrier penetration factor. In the fit, this line shape was fixed assuming decay from the 2$^+_2$ state only, with the Breit-Wigner parameters set to the tabulated values for this state and the penetration factor calculated for $\ell$=1 with a channel radius of $a$=4.85~fm. This allows for a good representation of the low-energy data in Fig.~\ref{fig:pp12C}(b), but we cannot rule out contributions  from the proposed doublet partner of the 2$^{+}_2$ state.   

The fitted yields for all decay branches are listed in Table~\ref{tab:level}. For the lowest-energy peak, we list only the yield up to $E^*$=8.3~MeV as our approximate line shape has a very long high-energy tail and much of the total yield is associated with this. In the fit,
 the structure above $E^*$=10.3~MeV is not entirely clear. We have included a moderately narrow peak at $E^*\approx$11.2~MeV to account for the rapidly drop off in yield at this energy. In addition we have added a very wide state between this and the 2$^+_3$ level. However its parameters are not well constrained and similar fits could be obtained  by replacing this wide level with some overlapping narrower peaks.

\subsubsection{Parent velocity distribution}  
\label{sec:2pCVcm}
The reconstructed $^{14}$O$^*$ velocity distributions for the 9.78 and 8.79 peaks are plotted in Fig.~\ref{fig:Vcm} as the dashed curves. The 9.78-MeV velocity distribution is sharply peaked at the $^{15}$O beam velocity similar to the 2$^+_1$ level and the doublet at the 2$^+_2$ energy, suggesting that is it also produced from the knockout of a valence $p$-shell neutron and hence is restricted to $J^{\pi}$=0$^+$, 1$^+$, or $2^+$.
 This is consistent with its tentative assignment of $J^{\pi}$=2$^+$ \cite{Solove:1991,Negret:2006}. The 8.79-MeV distribution has a broader peak in this velocity region and it is more similar in shape to the 1$^-_1$ peak in the $p$+$^{13}$N channel, possibly from the knockout of a $s_{1/2}$ neutron or some more complex process.

\subsubsection{Distributions of relative angle between protons}
  
\begin{figure}
\includegraphics[scale=.43]{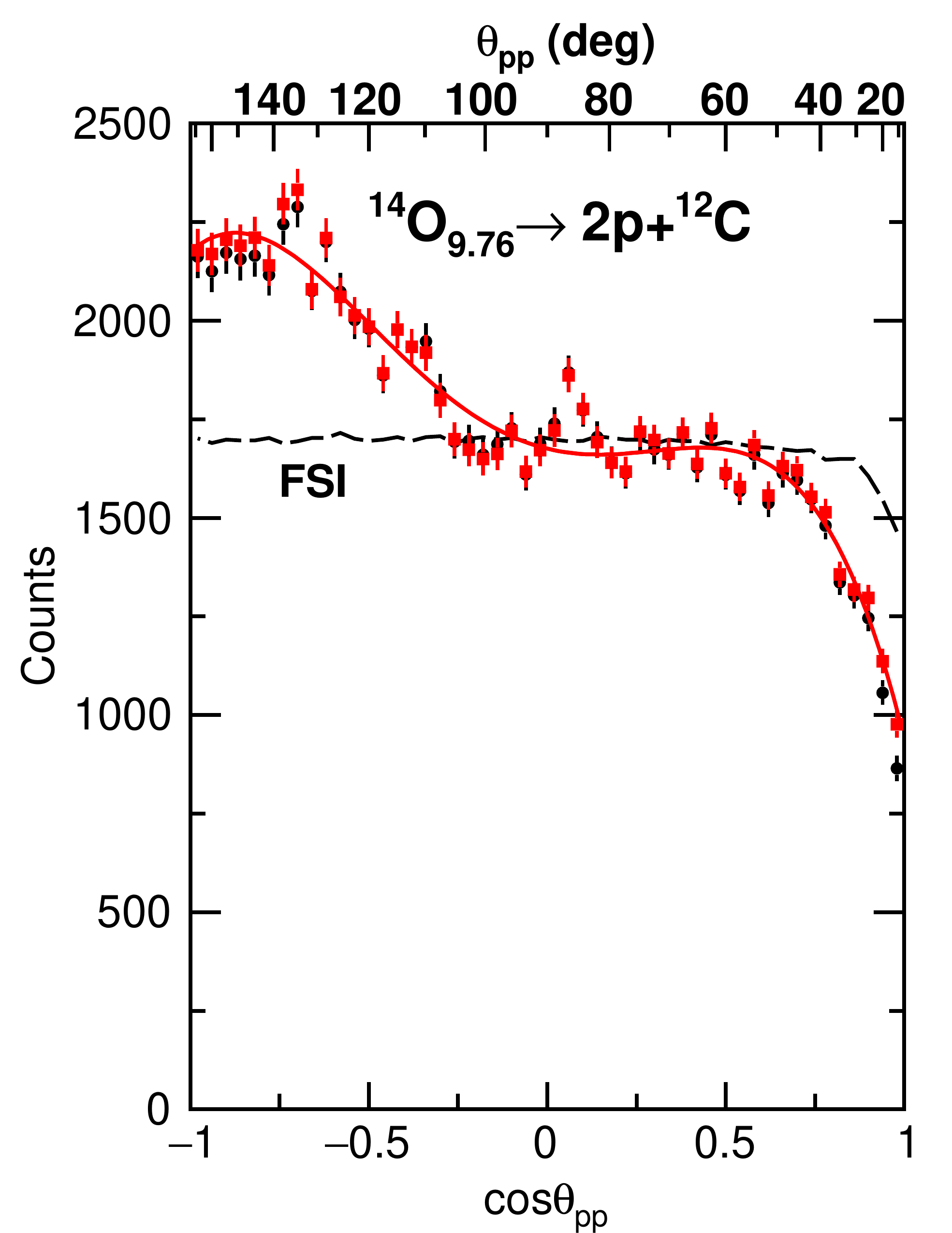}
\caption{The black circular data points give the distribution of relative-angle $\theta_{pp}$ between the two emitted 
protons in the sequential decay of the 9.78-MeV state. The dashed curves shows a calculation for isotropic proton emission including the Coulomb final-state  interaction between the protons. The red square data points have been corrected to remove this final-state interaction and also the effects of the detector resolution. The solid curve is a fit with a quartic function in $\cos\theta_{pp}$.}
\label{fig:thetapp}
\end{figure}

Information about the spin and parity of a sequential 2$p$ emitter can be gleaned from the distributions of relative angles $\theta_{pp}$ between the protons \cite{Charity:2009}. The $\theta_{pp}$ distribution for the 9.78~MeV state from the $^{15}$O beam obtained by gating only on the intense ridge in Fig.~\ref{fig:Jac}(c) ($^{13}$N intermediate state is either the 3/2$^-$ or 5/2$^+$) is plotted in Fig.~\ref{fig:thetapp} as the black circular data points.  No gate on transversely-emitted cores is applied here as its affect on the resolution of the relative angle is minor. This distribution has been corrected for the experimental acceptance as determined by our Monte Carlo simulations. In principle  $\theta_{pp}$ is just 180$^{\circ}$-$\theta_{k}$ when $\theta_k$ in Fig.~\ref{fig:jacy} is calculated when proton ``2'' is the second emitted proton. However in this case, we cannot distinguish which is the first and second emitted protons as they have similar kinetic energies. Therefore both ways of calculating $\theta_k$ are included in Fig.~\ref{fig:thetapp}. This has little effect as these two ways of calculating $\theta_k$ give very similar values. To further investigate this, and the effect of the detector resolution, the fitted distribution (as described below) is introduced into the Monte Carlo simulations and the simulated events analyzed in the same manor as the experimental data. The distribution obtained from this procedure was found to deviate by, at most, 3\%  from the 
primary distribution used as input into the simulations. This small deviation was used as a correction to generate the red square data points.

Classically this distribution should be symmetric about $\theta_{pp}$=90$^{\circ}$ ($\cos\theta_{pp}$=0). If the orbital angular momentum vectors of the two protons are aligned, then we expect a ``U'' shaped distribution with a minimum at $\theta_{pp}$=90$^{\circ}$, whereas if these vectors are perpendicular then the distribution will peak at 90$^{\circ}$. If one or both of the protons are emitted isotropically, or the two spin vectors have no angular correlations, then the distribution should be flat. A full quantum treatment can be found in Refs.~\cite{Frauenfelder:1953,Biendenham:1953}.  The experimental distribution does not follow any of these three possibilities and is quite remarkable having a distinct  asymmetry about $\theta_{pp}$=90$^{\circ}$ ($\cos\theta_{pp}$=0).  

There are two ways to introduce an asymmetry about $\theta_{pp}$=90$^{\circ}$. First, there is the possibility of final-state interactions (FSI) between the two protons. A sizable effect was observed for the 2$p$ decay of the 2\textsuperscript{nd} excited-state of $^{17}$Ne \cite{Charity:2018} where the second proton is emitted with a larger velocity than the first and thus, if emitted in the same direction, can catch up. Coulomb final-state interaction between the two protons resulted in a depletion of events near $\theta_{pp}$=0 compared to $\theta_{pp}$=180$^{\circ}$. From fitting the magnitude of this effect, the intrinsic width of $^{16}$F$_{g.s}$ could be extracted and was found to be consistent with the value obtained from directly measuring the width of the peak line shape \cite{Charity:2018}.  Here, the decay width of the two possible intermediate states are larger (62 and 47~keV)  than the value of 23~keV for the $^{16}$F$_{g.s.}$. However both protons are emitted with almost the same velocity so the second proton will not catch up to the first. To estimate the  magnitude of these interactions, we have used the classical calculation described in \cite{Charity:2018}. The predicted $\theta_{pp}$ distributions, assuming each proton is emitted isotropically is shown as the dashed black curve in Fig.~\ref{fig:thetapp} for decay through  the 5/2$^+$ intermediate state. The result for the 3/2$^-$ intermediate state is not that different. The magnitude of the final-state interaction is 
not large enough to explain the observed asymmetry. The red data points in Fig.~\ref{fig:thetapp} also contain a correction to remove the effects of the final-state interactions based on this calculation. This is largely for display purposes as subsequently we do not fit the region of small $\theta_{pp}$ ($\cos\theta_{pp}>$0.8) where the effect of the FSI are significant 

The second way to produce the asymmetry relies on the unusual occurrence in this decay that there are two possible intermediate states in $^{13}$N which are degenerate (3/2$^-$ and 5/2$^+$) and thus we cannot distinguish energetically whether the decay is through only one of these or there are decay branches though both of them. If the latter is true, then there can be interference between these two decay paths. In addition as the parity of the two degenerate states are opposite, this interference term will be asymmetric. In general if there are two decay paths ``a'' and ``b'' with amplitudes $\alpha$ and $\beta$, then the relative angular distribution will be \cite{Biendenham:1953}. 
\begin{equation}
W(\theta_{pp}) = \lvert \alpha \rvert^2 W_{a}(\theta_{pp}) + \lvert \beta \rvert^2 W_{b}(\theta_{pp})  + (\alpha \beta^* + \alpha^* \beta) W_{ab}(\theta_{pp})
\end{equation} 
where $W_a(\theta_{pp})$ and $W_{b}(\theta_{pp})$ are the distributions for  decay purely through the ``a'' and ``b'' paths, respectively, and $W_{ab}(\theta_{pp})$ is the interference term. Normally one considers a parent state with  mixed a configuration that has decay paths through the same intermediate state. For example if the spin of this state is 2$^+$, then we can consider  both  $s_{1/2}$ and $d_{5/2}$ decays to the 5/2$^+$ intermediate state. The interference term is such cases will be symmetric about $\theta_{pp}$=90$^{\circ}$.   However the interference between a $p_{1/2}$ decay to the 3/2$^-$ state and a $s_{1/2}$ decay to the 5/2$^+$ state  will be asymmetric.  We have considered the possibility of  $p_{1/2}$ and $p_{3/2}$ decays to the 3/2$^-$ intermediate state and $s_{1/2}$, $d_{3/2}$, and $d_{5/2}$ decays to the 5/2$^+$ intermediate state. This gives 10 possible interference terms, six of which are asymmetric. The asymmetric terms can either enhance or suppress the small relative angles depending on the relative phase of their two components. With all these contributions, the final relative-angle distribution is quartic in $\cos\theta_{pp}$. The best quartic fit is shown as the solid red curve in Fig.~\ref{fig:thetapp} which reproduces the experimental data quite well.

 With five possible decay branches and their phases we have nine parameters (one of the phases can be set to zero as only relative phases are important).
However as the final distribution is a quartic function of $\cos\theta_{pp}$ with only five parameters, it is not surprising that our best fit can be reproduced by many different sets of branching ratios and phases. There are however some restrictions on the branching ratios. The presence of a $\cos\theta_{pp}^{4}$ term requires a decay branch where the two emitted  protons have  $d_{5/2}$ character or $\ell\geq$3. We will ignore the latter possibility. With the presence of this $d_{5/2}$-$d_{5/2}$ decay path, we obtain $\cos^4\theta_{pp}$ terms from the distribution for pure $d_{5/2}$-$d_{5/2}$ decay and from  its interference with the $d_{3/2}$-$d_{5/2}$ decay path. The magnitude of the fitted $\cos^4\theta_{pp}$ coefficient  constrains the branching ratio by $d$-wave emission to the 5/2$^+$ state in $^{13}$N to be at least 13.2\% of the total branching ratio to the doublet.

Of course there must be decay paths to both members of the doublet in order to get the asymmetric interference terms. Consider two extremes. Firstly, if the decay to the 5/2$^+$ intermediate state dominates, then we find there must be at least a 3.8\% decay branch to the 3/2$^-$ state. On the other hand if decay to the 3/2$^-$ state dominates, then we require at least a 15.5\% decay branch to the 5/2$^+$ intermediate state. Clearly it only takes a small admixture of one of the decay branches to the other to produce a significant asymmetric distribution. Therefore our observation of this asymmetry is not surprising.  



\section{Theoretical results}
\label{sec:theory}

To describe spectra of $^{14}$O resonances and their properties, we used the Shell Model Embedded in the Continuum (SMEC) which describes resonances in the framework of the shell model for open quantum systems  \cite{smec,smec1,smec11,smec2,smec3}. Recent SMEC applications included the study of correlations and clustering in near-threshold states \cite{smec2,smec3,oko2016,oko2018} and the change of the continuum coupling strength for like and unlike nucleons in exotic nuclei \cite{Charity:2018}. 

Here, the scattering environment is provided by the one-proton decay channels. The Shell Model (SM) states are coupled to the environment of one-proton decay channels through the energy-dependent effective Hamiltonian:
\begin{equation}
{\cal H}(E)=H_{{\cal Q}_0{\cal Q}_0}+W_{{\cal Q}_0{\cal Q}_0}(E),
\label{eq21}
\end{equation}
where ${\cal Q}_{0}$ and ${\cal Q}_{1}$ denote the two orthogonal subspaces of Hilbert space containing 0 and 1 particle in the scattering continuum, respectively. $H_{{\cal Q}_0{\cal Q}_0}$ stands for the standard SM Hamiltonian and $W_{{\cal Q}_0{\cal Q}_0}(E)$:
\begin{equation}
W_{{\cal Q}_0{\cal Q}_0}(E)=H_{{\cal Q}_0{\cal Q}_1}G_{{\cal Q}_1}^{(+)}(E)H_{{\cal Q}_1{\cal Q}_0},
\label{eqop4}
\end{equation}
is the energy-dependent continuum coupling term, where $G_{{\cal Q}_1}^{(+)}(E)$ is the one-nucleon Green's function and 
${H}_{{Q}_0,{Q}_1}$, ${H}_{{Q}_1{Q}_0}$ are the coupling terms between the subspaces ${\cal Q}_{0}$ and ${\cal Q}_{1}$. $E$ stands for a scattering energy of particle in the continuum and the energy scale is defined by the lowest one-proton decay threshold. Each decay-channel state in $^{14}$O is defined by the coupling of one proton in the scattering continuum of $^{13}$N in a given SM state. Internal dynamics of the  ${\cal Q}_{0}$  system include couplings to the environment of decay channels and is given by the energy-dependent effective Hamiltonian ${\cal H}(E)$.

Each decay threshold is associated with a non-analytic point of the scattering matrix. The coupling of different SM eigenfunctions to the same decay channel induces a mixing among the SM eigenfunction which reflects the nature of the decay threshold.  This continuum-induced configuration mixing, which consists of the Hermitian principal-value integral describing virtual continuum excitations and the anti-Hermitian residuum that represents the irreversible decay out of the internal space ${\cal Q}_{0}$, can radically change the structure of near-threshold SM states. Unitarity in a multichannel system implies that the external mixing of SM eigenfunctions changes whenever a new channel opens up. Hence, the eigenfunctions depend on the energy window imposed theoretically, {\em i.e.} on the chosen set of decay channels, and vary with the increasing total energy of the system, {\em i.e.} with the variation of the environment of scattering states and decay channels.

The effective Hamiltonian ${\cal H}(E)$ is Hermitian for energies below the lowest particle-emission threshold and non-Hermitian above it. Consequently, the energy-dependent solutions of ${\cal H}(E)$ are real for bound states ($E<0$) and complex 
$\tilde{E}_i - (1/2) i \tilde{\Gamma}_i$ in the continuum ($E>0$). In general, due to the energy dependence of $\tilde{E}_i$
 and $\tilde{\Gamma}_i$, the line shape of the resonance differs from a Breit–Wigner shape, even without any interferences and far from the decay thresholds.

The energy and the width of resonance states are determined by the fixed-point conditions \cite{smec}: 
\begin{eqnarray}
E_i &=& {\tilde E}_i(E=E_i)   \nonumber  \\
\Gamma_i &=& {\tilde \Gamma}_i (E=E_i). 
\label{eqfp}
\end{eqnarray}
The solution of these equations sets the energy scale for states with the quantum numbers $J$, $\pi$ of state $i$. In practice, to have the same scale for the whole spectrum, an energy reference is defined by a single representative state, {\em e.g.} the ground state of the nucleus. However, to compare parameters of a physical resonance (widths, branching ratios) with experimental data, 
we solve the fixed-point equations (\ref{eqfp}) for each  resonance studied at  the experimental energy of this resonance relative to the elastic-channel threshold.

The SMEC Hamiltonian in this work consists of the monopole-based SM interaction (referred to as YSOX \cite{yuan2012}) in the full $psd$ model space plus the Wigner-Bartlett contact interaction \cite{Shalit:1963}: 
\begin{equation}
V_{12}=V_0 \left[ \alpha + \beta P^{\sigma}_{12} \right] \delta\langle\bf{r}_1-\bf{r}_2\rangle ~ \ ,
\label{WB}
\end{equation}
for the coupling between SM states and the decay channels, where $\alpha + \beta = 1$ and $P^{\sigma}_{12}$ is the spin exchange operator. The SM eigenstates have good isospin but the continuum-coupling term $W_{{\cal Q}_0{\cal Q}_0}(E)$ in ${\cal H}(E)$ breaks the isospin conservation due to different radial wave functions for protons and neutrons, and different proton and neutron separation energies.


\begin{table}[h!]
	\caption{\label{table_spectra} The widths of known $^{14}$O resonances  calculated in the SMEC using the modified YSOX interaction and  the Wigner-Bartlett continuum coupling with  strength $V_0 = -350$ MeV$\cdot$fm$^3$. Fixed-point equations (\ref{eqfp}) are solved for each resonance at energy $E$ equal to its experimental energy relative to the elastic-channel threshold. Comparsions to experimental width $\Gamma_{\rm exp}$ are made. 
		(For more details, see the text.) }
		\begin{ruledtabular}
		\begin{tabular}{cccc}
			$J^{\pi}$  & $E^{*}_{\rm exp}$ & $\Gamma_{\rm exp}$ & $\Gamma_{\rm th}$  \\ 
		\hline 
			$1^-_1$ & 5.164(12) & 38(2) & 57.5 \\
			$0^-_1$ & 5.71(14) & 400(44) & 280 \\
			$0^+_2$ & 5.931(10) & $<$12 & 22 \\
			$3^-_1$ & 6.285(12) & 37.7(17) & 655 \\
			$2^+_1$ & 6.585(11) & $<$25 & 0.1 \\
			$2^-_1$ & 6.764(10) & 96(5) & 964 \\
			$2^+_2$ & 7.768(10) &  68(6) & 8.39 \\
			$2^+_3$ & 9.755(10) & 229(51) & 563 \\
		        
					\end{tabular}
		\end{ruledtabular}
	
		\end{table}		
The radial single-particle wave functions (in ${\cal Q}_0$) and the scattering wave functions (in ${\cal Q}_1$) are generated by the Woods-Saxon (WS) potential which includes spin-orbit and Coulomb parts. The radius and diffuseness of the WS potential are $R_0=1.27 A^{1/3}$ fm and $a=0.67$ fm, respectively. The spin-orbit potential is $V_{\rm SO}=6.4$~MeV,  and the Coulomb part is calculated for a uniformly charged sphere with radius $R_0$.  The depth of the central part for protons is adjusted to reproduce the measured proton separation energy ($S_p$=4.627 MeV) for the  $p_{1/2}$ single-particle state . Similarly, the depth of the potential for neutrons is chosen to reproduce the measured neutron separation energy  ($S_n$=23.179 MeV) for the  $p_{3/2}$ single-particle state. 

The original YSOX interaction fails to reproduce energies of the low-lying unnatural parity states in $^{14}$O. Spectrum of these states was greatly improved if the  cross-shell $T=1$ matrix elements: $\langle 0p_{1/2}1s_{1/2};J^{\pi}=0^-|V|0p_{1/2}1s_{1/2};J^{\pi}=0^-\rangle$ and $\langle 0p_{1/2}1s_{1/2};J^{\pi}=0^-|V|0p_{1/2}1s_{1/2};J^{\pi}=1^-\rangle$  were made smaller by 1 MeV and 2 MeV, respectively. This change implies a modification of the $T=1$ monopole term ${\cal M}^{T=1}(0p_{1/2} 1s_{1/2})$ which becomes -1.067 MeV, as compared to +0.683 MeV in the original YSOX interaction. No change was made in the $T=0$ monopole, ${\cal M}^{T=0}(0p_{1/2} 1s_{1/2}) = -2.842$ MeV.
The continuum-coupling constant of the Wigner-Bartlett interaction $V_0 = -350$ MeV$\cdot$fm$^3$  has been chosen to obtain an overall satisfactory reproduction of the spectrum of proton resonances in $^{14}$O. The spin-exchange parameter $\alpha$ ($\beta=1-\alpha$) in the coupling between SM states and the decay channels [Eq.~(\ref{WB})] has a standard value of $\alpha = 0.73$ \cite{smec}.
 
The SMEC results are compared with the experimental excitation energies in Fig. \ref{fig:level}.
In this calculation, for each $J^{\pi}$ separately, the SM states are mixed via the coupling to 9 channels, including the elastic channel $[{^{13}}$N($1/2^-) \otimes {\rm p}(\ell_j)]^{J^{\pi}}$ and 8 inelastic channels $[{^{13}}$N($K^{\pi}) \otimes {\rm p}(\ell_j)]^{J^{\pi}}$ which correspond to the excited states of $^{13}$N: $K^{\pi}=1/2^+_1$, $3/2^-_1$, $5/2^+_1$, $5/2^+_2$, $3/2^+_1$, $7/2^+_1$, $5/2^-_1$, and $3/2^+_2$. 
The fixed-point equation in the SMEC calculation is solved for the ground state $0^+_1$. With this choice, the SMEC and experimental ground-state energy have the same origin of the energy scale. The agreement between experimental and calculated spectrum is quite remarkable. The one conspicuous failure of these calculations is the over prediction of the  one-proton emission widths of the higher-spin unnatural-parity resonances, $3^-_1$ and $2^-_1$, (see Table \ref{table_spectra}). 

The experimental results suggest that the $2^+_2$ resonance coincides with another resonance of unknown spin and parity. This resonance doublet decays mainly to the $1/2^-$ ground state  $^{13}$N with a small decay branch to the first-excited state $1/2^+$. The SMEC predicts that the partner of the $2^+_2$ resonance could be $0^+_3$ resonance. This assignment is also favored by the observed parent velocity distribution of the doublet (see Secs.~\ref{sec:pCVcm} and \ref{sec:2pCVcm}).    

\begin{figure}[t!]
	\includegraphics[width=1.00\linewidth]{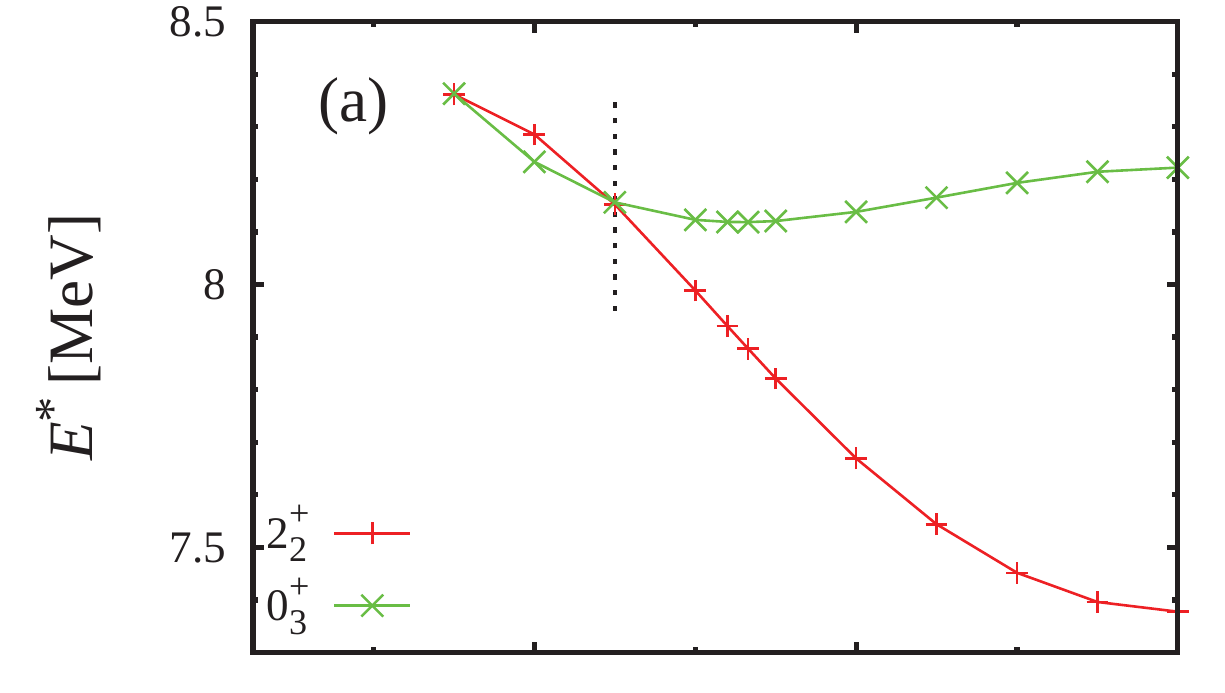}\\
	\vskip -.1truecm
	\includegraphics[width=1.0\linewidth]{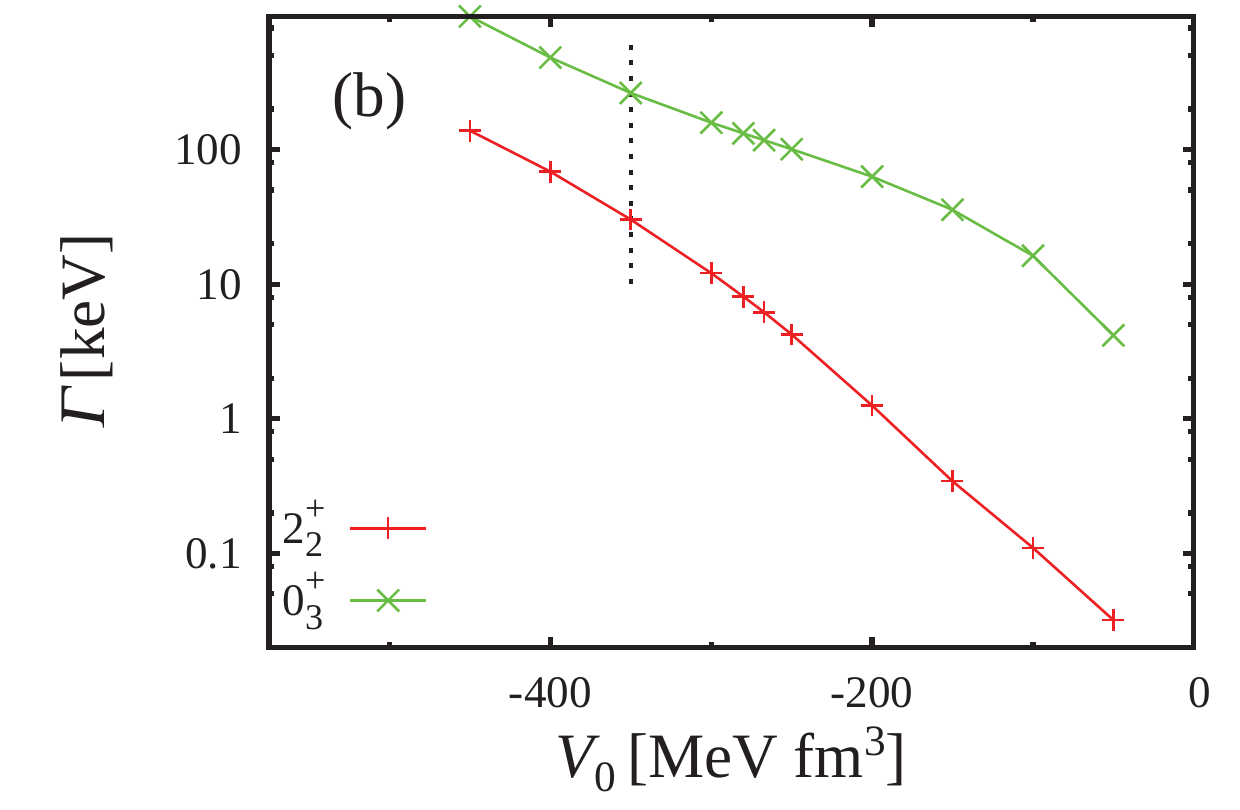}
\caption{ Results for the $2^+_2$ and $0^+_3$ levels in the SMEC as a function of the  continuum coupling strength $V_0$. Panel (a) shows the eigenenergies and  panel (b)  the widths of SMEC eigenstates which form a doublet of resonances in the interval 
$-450 \leq V_0 \leq -330$ MeV$\cdot$fm$^3$. 
The dotted vertical line shows the value of $V_0 = -350$ MeV$\cdot$fm$^3$ which gives an overall reasonable reproduction of the resonances reported in this work, see Fig.~\ref{fig:level}}
	\label{fig_1add}
\end{figure}

The dependence of the $2^+_2$ and $0^+_3$ eigenvalues on the continuum-coupling strength is shown in Fig. \ref{fig_1add}. 
Below $V_0 \approx -330$ MeV$\cdot$fm$^3$ these eigenvalues are nearly degenerate [panel (a)].
The calculated (total) decay widths of the $2^+_2$ and $0^+_3$ resonances are changing smoothly with increasing continuum coupling [panel (b)] and at 
$V_0 = -350$ MeV$\cdot$fm$^3$ are 8.4 keV and 129 keV, respectively. The branching ratios exhibit a weak dependence on the continuum-coupling strength. For $V_0 = -350$ MeV$\cdot$fm$^3$, $\Gamma(2^+_2 \rightarrow 1/2^-)/\Gamma(\rm tot)= 0.25$, whereas 
$\Gamma(0^+_3 \rightarrow 1/2^-)/\Gamma(\rm tot)= 0.73$. Hence, the narrow $2^+_2$ resonance is predicted to decay mainly to the first-excited state $1/2^+$, while the broader $0^+_3$ resonance decays predominantly to the ground state of $^{13}$N.

The present work confirms the existence of a resonance at  8.787(13) MeV with a width of $\Gamma = 182(32)$~keV which is listed as tentative in the most recent  evaluation.
This resonance decays both to the $3/2^-_1$, $5/2^+_1$ doublet and to the $1/2^+_1$ resonance (more to the former than the latter), and while the peak associated with decay to the ground state was not observed, we place an upper limit of 65\% for this branch (Table~\ref{tab:level}). 
The SMEC calculations predict $1^-_2$, $2^-_2$, and $3^+_1$ resonances in this energy region which have not been connected to experimental levels. 
Their predicted decay widths are 200 keV,  87 keV, and 744 keV, respectively. 
The predicted ground-state branch for the $2^-_2$ resonance  (81\%) largely exceeds the experimental upper limit.
The predicted branching fractions for the $1^-_2$ resonance to the ground state, $1/2^+_1$ first-excited state and to the $3/2^-_1$, $5/2^+_1$ doublet are:
50\%, 9\% and 41\%, respectively. Finally, the $3^+_1$ resonance is predicted to decay mainly to $1/2^+_1$ first-excited state (89\%) and the remaining fraction goes to the $3/2^-_1$, $5/2^+_1$ doublet. As compared to the experiment, this resonance has both inverted decay fractions and significantly larger total decay width. Thus the predicted energy, width and branching fractions for the $1^-_2$ state are consistent with those for the 8.787 MeV resonance. As this cannot be said for the other two candidates, this  assignment is made. 

A resonance at 11.195(30) MeV, with a width $\Gamma < 220$ keV was observed to decay to the $3/2^-_1$, $5/2^+_1$ pair. 
No peak associated with decay to the ground state was observed, however again due to poor efficiency for this decay, we can only constrain this branching ratio to $<$81\% (see Table~\ref{tab:level}).
In the vicinity of this resonance, the SMEC calculations predict $2^+_4$, $3^-_2$, and $3^+_2$ resonances 
with total decay width $\Gamma(\rm tot)$ equal 268 keV, 1700 keV, and 103 keV, respectively.  
The  $2^+_4$ state is predicted to decay mainly to the $3/2^-_1$, $5/2^+_1$ doublet with a branching ratio: $\Gamma(2^+_4 \rightarrow 1/2^-)/\Gamma(\rm tot) = 0.7$. The remaining flux goes to the $1/2^+_1$ first-excited state. 
The $3^-_2$ resonance is predicted to decay to the doublet with a probability 98\% (the remaining flux in this decay goes to the ground state) but its very-large predicted width removes it as a suitable candidate. 
Finally, the $3^+_2$ resonance decays mainly to the first-excited state $1/2^+_1$ with a probability 60\% and the remaining flux goes to the doublet, {\em i.e.} branching fractions inverted from what is observed.
For the observed resonance at 11.195(30) MeV, $2^+_4$  is the most suitable assignment (best overall agreement with energy, decay width, branching ratios), however $3^+_2$ cannot be excluded because despite its inverted decay fractions, it has a suitable decay width. As the present calculations do not include coupling to the $\alpha$+$^{10}$C channel the above assignment is tentative.

\subsection{Configuration mixing in resonance spectra}

Continuum induced mixing of SM eigenstates is strong if there are  avoided crossings of SMEC eigenstates \cite{smec,oko2009}. These crossings can be conveniently studied by calculating energy trajectories of the double poles of the scattering matrix, the so-called exceptional points (EPs) \cite{zirn83}, of the effective Hamiltonian with the complex-extended continuum coupling strength $V_0$. The connection of EPs to avoided crossings, spectral properties \cite{hei91,Dukelsky:2009}, and associated geometric phases have been discussed in simple models in considerable detail \cite{heis98,heis00,Dembowski:2001,Dembowski:2003,Keck:2003}. Their manifestation in scattering experiments has been considered in \cite{oko2009}. Since the effective Hamiltonian is energy dependent, the essential information about configuration mixing is contained in the trajectories of coalescing eigenvalues $E_{i_1}(E) = E_{i_2}(E)$, the so-called exceptional threads (ETs), for a complex value of the continuum coupling strength $V_0$.
 
EPs correspond to common roots of the equations:
\begin{equation}
 \frac{\partial^{(\nu)}}{\partial {\cal E}} {\rm det}\left[{\cal H}\left(E;V_0\right)  -{\cal E}I\right] = 0,~~~\nu=0,1.
\label{discr}
\end{equation}

Single-root solutions of Eq. (\ref{discr})  correspond to EPs associated with either decaying or capturing states. The maximum number of such roots is:
$M_{\rm max} = 2n(n - 1)$, where $n$ is the number of states of given angular momentum $J$ and parity $\pi$. The factor 2 in the expression on $M_{\rm max}$ comes from the symmetry of solutions of ${\cal H}(E)$ with respect to the transformation $V_0 \rightarrow -V_0$. This symmetry is broken above the lowest particle-emission threshold but it rarely happens that the ET change from a physical value $V_0<0$ to an unphysical value for $V_0$. 

Below the first particle-emission threshold, half of all ETs correspond to the poles with the asymptotic of a decaying state whereas the other half has the capturing state asymptotic. In the continuum, this symmetry is broken, and at higher excitation energies, the analytic continuation of a decaying pole in the coupling constant may become the capturing pole, or {\em vice versa}. It should be stressed that both decaying and capturing poles influence the configuration mixing in physical SMEC wave functions.

It has been shown \cite{smec2,smec3} that ETs exhibit generic features that are fairly independent of both the continuum-coupling strength and the detailed nature of the coupling matrix elements. This makes ETs particularly suitable for the investigation of the susceptibility of the eigenfunctions to the features of the coupled multichannel network  in the whole domain of the continuum-coupling strength.

\subsubsection{$2^+$ resonances}

\begin{figure}[t!]
	\includegraphics[width=1.00\linewidth]{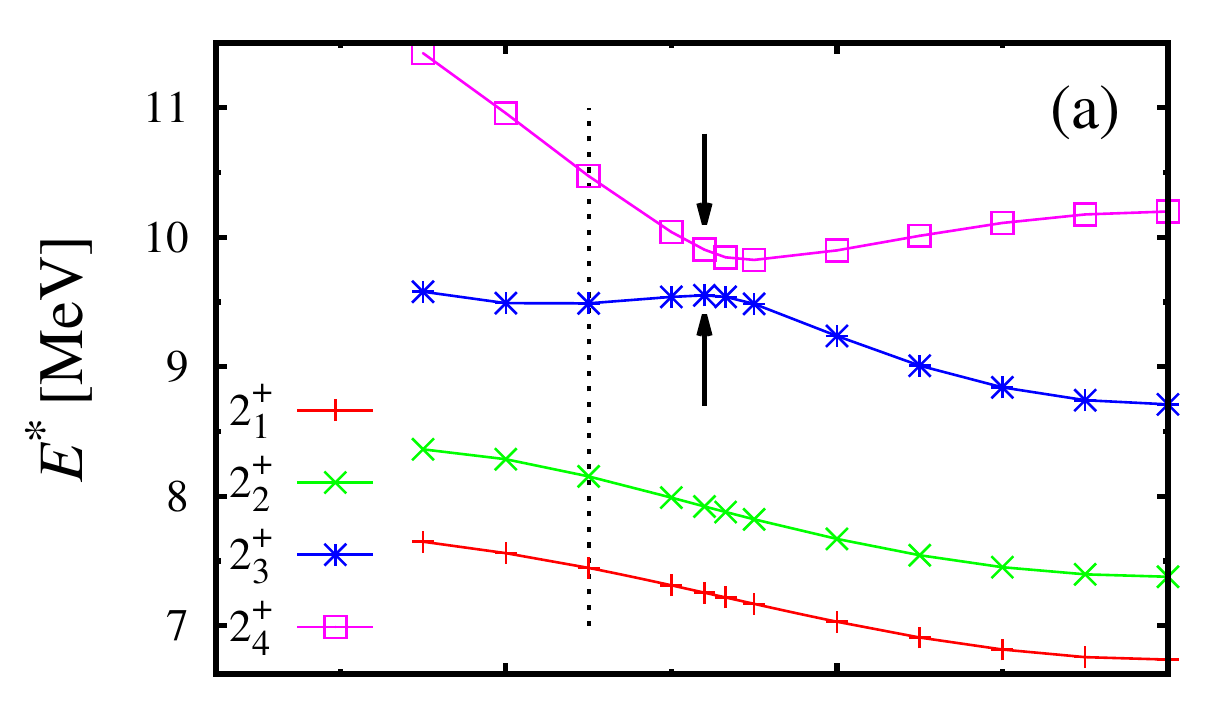}\\
	\vskip -.4truecm
	\includegraphics[width=1.00\linewidth]{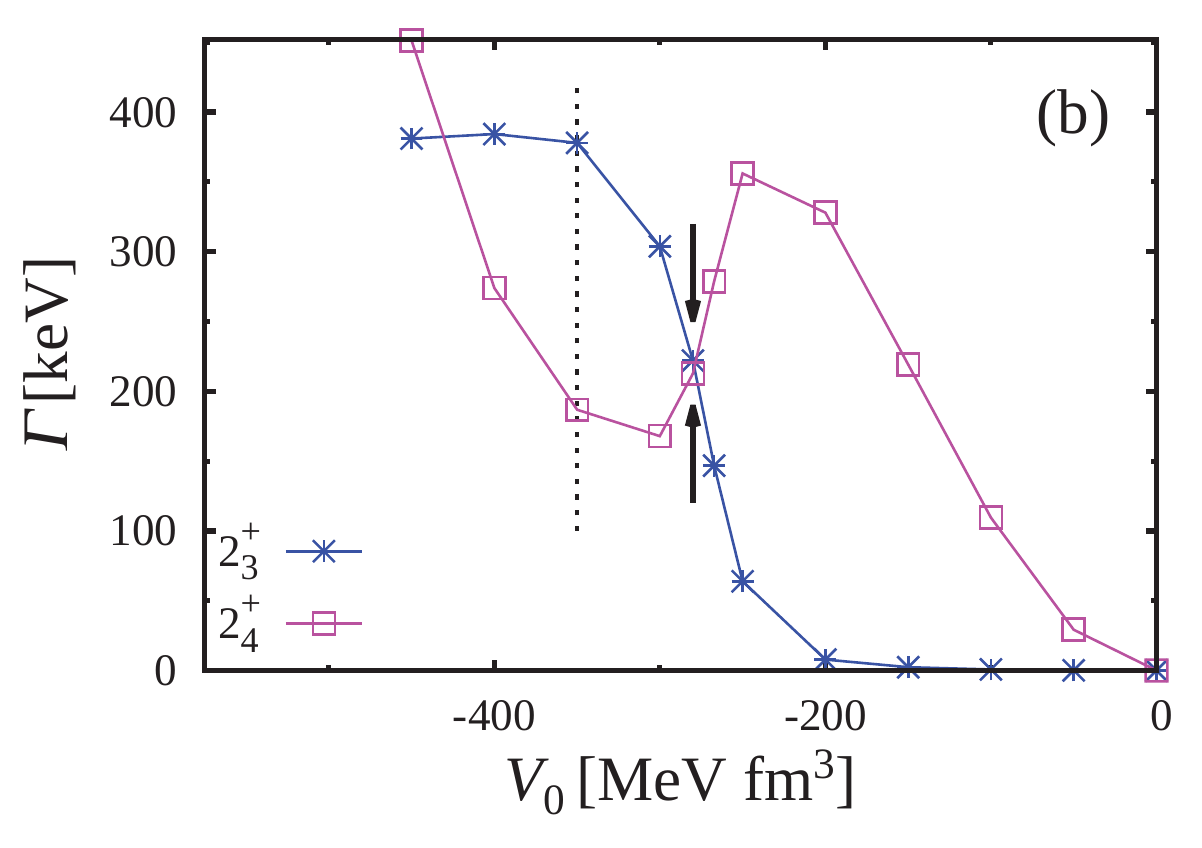}
\caption{The SMEC spectrum of $2^+_i$ eigenstates ($i=1,2,3,4$) in $^{14}$O is shown as a function of the (real) continuum coupling strength $V_0$.  Panel (a) presents the eigenenergies and  panel (b) the widths of $2^+_3$  and $2^+_4$ SMEC eigenstates, the states that are involved in the avoided crossing at $V_0 \approx -280$ MeV$\cdot$fm$^3$, see vertical arrows.  The dotted vertical line shows the value of $V_0 = -350$ MeV$\cdot$fm$^3$ which gives an overall reasonable reproduction of the resonances reported in this work. For more details, see the text.}
	\label{fig_2a}
\end{figure}

The upper panel of Fig. \ref{fig_2a} shows the spectrum of four $2^+$ eigenvalues of the SMEC Hamiltonian  plotted as a function of the (real) continuum coupling constant $V_0$. For all $V_0$, the fixed-point equation is solved at the ground-state $0^+_1$. The $2^+_1$ and $2^+_2$ eigenenergies decrease monotonically with increasing $V_0$. However the $2^+_3$ and $2^+_4$ eigenvalues have an avoided crossing near $V_0 \approx -280$ MeV$\cdot$fm$^3$ corresponding to  $E^* \approx 9.7$ MeV. 

The evolution of the $2^+_3$ and $2^+_4$ resonance widths as a function of $V_0$ is shown in the lower panel of Fig. \ref{fig_2a}. For lower values of $|V_0|$, one can see a significantly different dependence of the $2^+_3$ and $2^+_4$ eigenvalues to the continuum coupling. With increasing  $|V_0|$, at first the width of $2^+_4$ resonance is both much larger and grows much faster than the width of $2^+_3$ resonance until $V_0\approx -250$ MeV$\cdot$fm$^3$.
The width of the $2^+_3$ resonance then decreases and the two widths coincide for $V_0 \approx -280$ MeV$\cdot$fm$^3$.
This indicates that the $2^+_3$ and $2^+_4$ eigenvalues are close to a $2^+$ EP. For even more negative coupling strengths, the width of higher energy $2^+_4$ resonance again increases rapidly overtaking that of the  $2^+_3$ resonance at $V_0 \approx -425$ MeV$\cdot$fm$^3$.

\begin{figure}[t!]
	\includegraphics[width=1.00\linewidth]{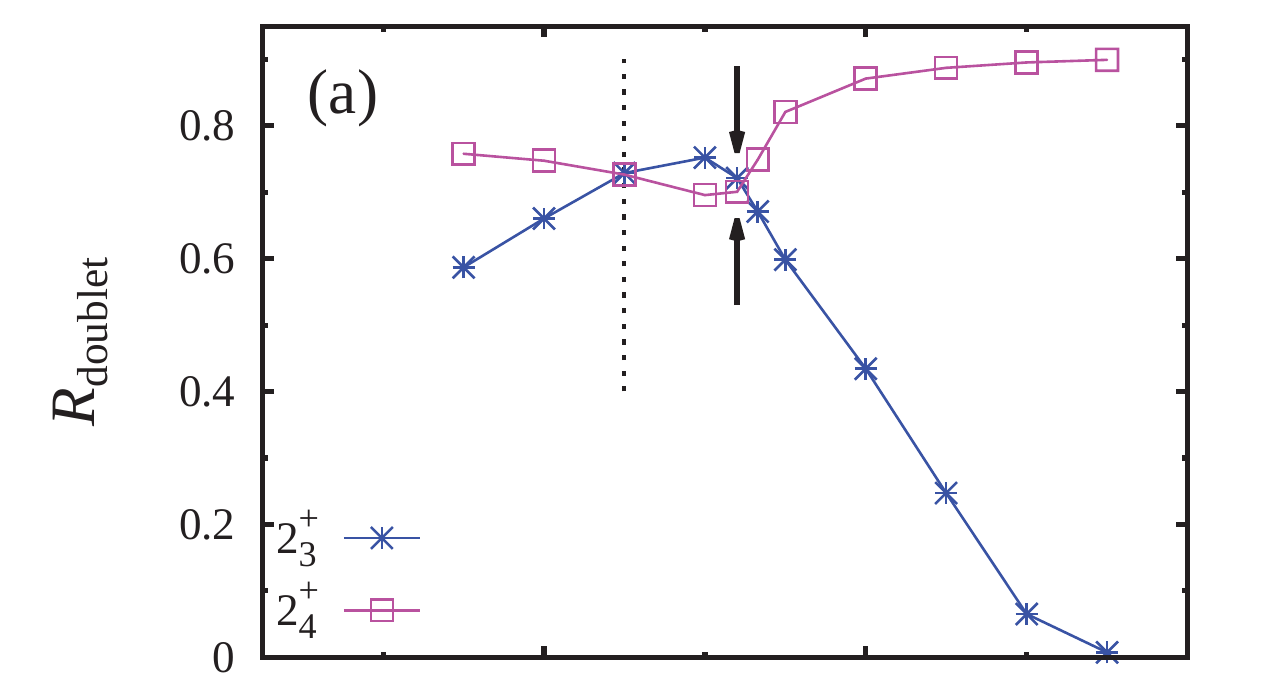}\\
	\vskip -.3truecm
 
	\includegraphics[width=1.00\linewidth]{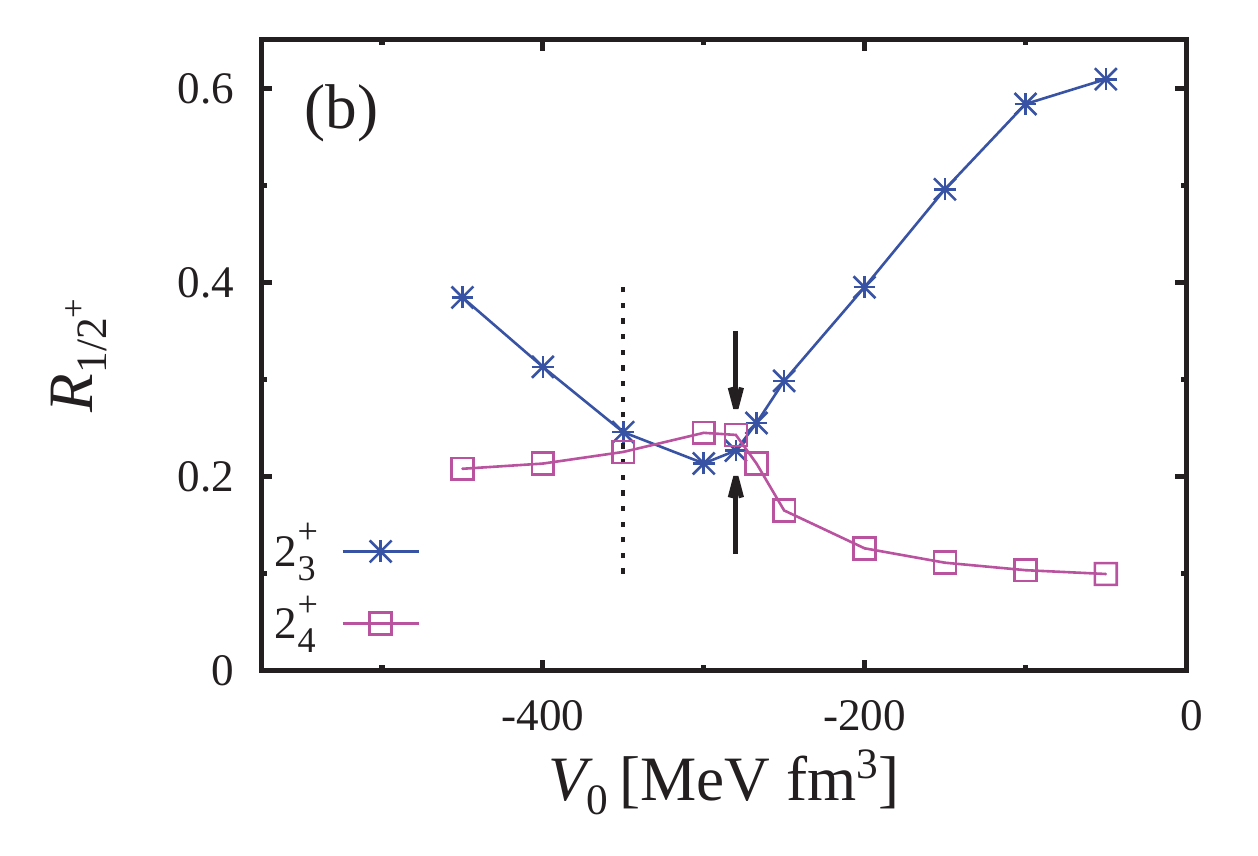}
\caption{The branching fractions, as a function of the continuum-coupling strength $V_0$, for the one-proton decay of $2^+_3$ and $2^+_4$ resonances either to (a) the doublet of resonances  $3/2^-_1$ - $5/2^+_1$ or (b) the first-excited state $1/2^+_1$ in $^{13}$N. The arrows show the location of the avoided crossing and the dotted vertical line is at $V_0 = -350$ MeV$\cdot$fm$^3$, the value of the $V_0$ that gives a satisfactory reproduction of the observed $^{14}$O spectrum.}
	\label{fig_2ab}
\end{figure}

Figure \ref{fig_2ab} shows the behavior of partial decay rates of the $2^+_3$ and $2^+_4$ resonances.  For small values of $|V_0|$, the decay properties of $2^+_3$ and $2^+_4$ eigenvalues are quite different: the $2^+_3$ resonance decays mainly to the first-excited state $1/2^+_1$ [panel (b)] whereas $2^+_4$ decays to the $3/2^-_1$ - $5/2^+_1$ doublet [panel (a)]. The decay fractions remain approximately the same for the two resonances from the avoided crossing ($V_0 \approx -280$ MeV$\cdot$fm$^3$) until $V_0 \simeq -350$ MeV$\cdot$fm$^3$ and then diverge for even larger $|V_0|$.

\begin{figure}[t!]
	\includegraphics[width=1.00\linewidth]{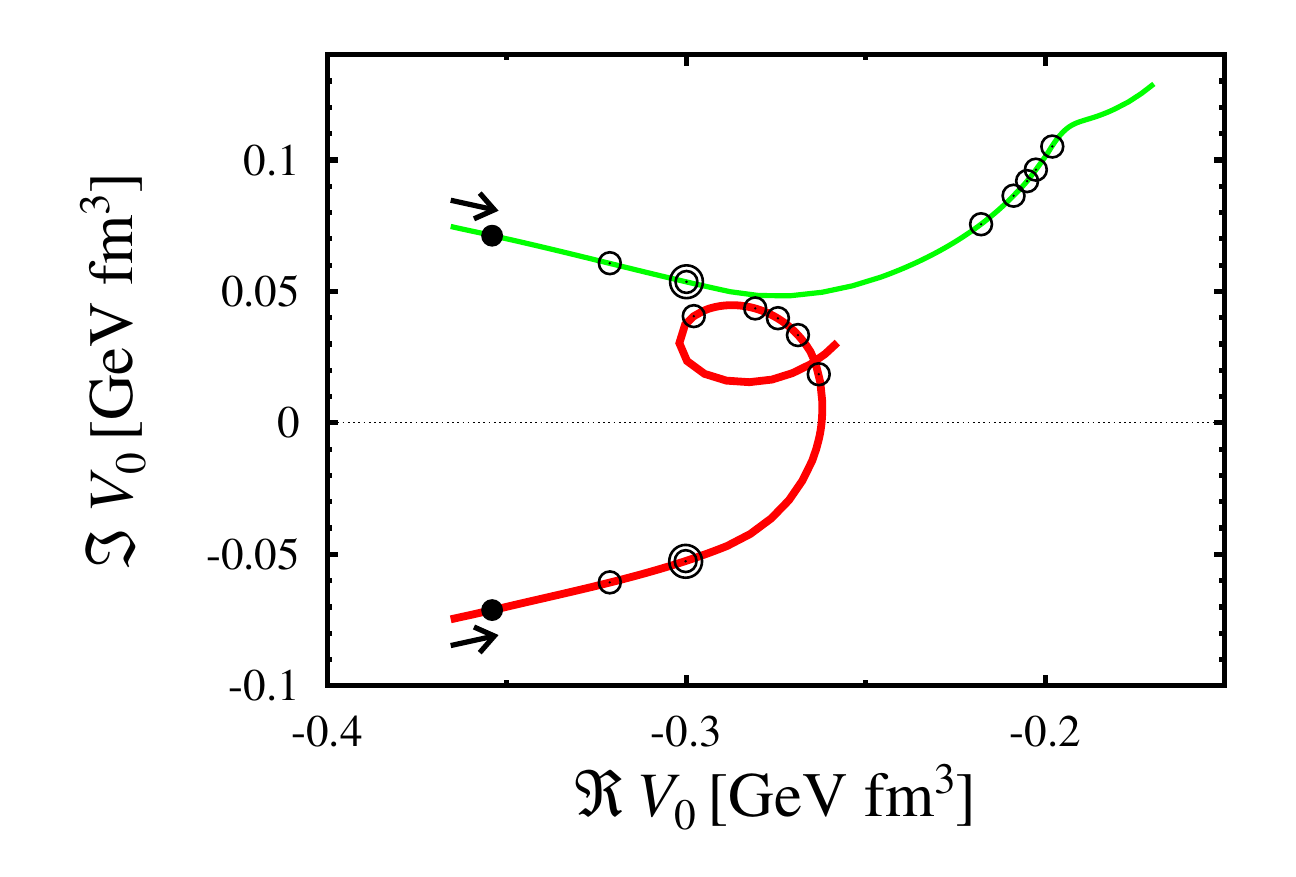}\\
\caption{Exceptional threads for $2^+$ eigenstates of SMEC in $^{14}$O are shown as a function of the real and imaginary parts of the continuum coupling strength $V_0$. Negative (positive) imaginary values of $V_0$ correspond to outgoing, {\em i.e.} decaying, (ingoing, {\em i.e.} capturing) asymptotics.
Different points on these ETs correspond to different excitation energies. The filled circle points to the threshold energy of the lowest decay channel (the elastic reaction channel). Open circles denote excitation energies at which subsequent (inelastic) channels open. The open double circle corresponds to the opening of two nearly-degenerate decay channels at 8.129 MeV and 8.174 MeV. Arrows indicate the direction of increasing excitation energy along each ET.}
	\label{fig_2b}
\end{figure}

Figure \ref{fig_2b} displays two exceptional threads (ETs) in the complex plane ${\cal R}e(V_0)$ - ${\cal I}m(V_0)$ that are both in the physical region, close to the real axis, and in the energy region relevant for the mixing of the $2^+_3$ and $2^+_4$ resonances. The ET shown by the thick red curve, which crosses the real axis at $V_0 = -262$ MeV$\cdot$fm$^3$, is responsible for the avoided crossing (see Fig. \ref{fig_2a}).
It should be noted that in this presentation,  the ET dynamics is independent of the reference state chosen for solving the fixed-point equation (\ref{eqfp}). 

Each point along the ET corresponds to a different scattering energy $E$.
For energies below the lowest proton decay threshold 
(the elastic-channel threshold: $[{^{13}}$N($1/2^-_1) \otimes {\rm p}(\ell_j)]^{J^{\pi}}$)
both ETs shown in Fig. \ref{fig_2b} are straight lines with reflection symmetry with respect to the real axis. 
This symmetry is broken above the first decay threshold. Different circles show the points (energies) at which various one-proton decay channels open. 
The filled circle denotes the lowest-energy decay threshold. The open double circle shows the nearly degenerate second and third decay channels, 
$[{^{13}}$N($3/2^{-}_1) \otimes {\rm p}(\ell_j)]^{J^{\pi}}$ and $[{^{13}}$N($5/2^{+}_1) \otimes {\rm p}(\ell_j)]^{J^{\pi}}$,
which open at 8.129 MeV and 8.174 MeV, respectively.

The ET depicted by the thick red curve, which at low excitation energies corresponds to the succession of double poles of the $S$ matrix with the decaying asymptotic [${\cal I}m(V_0) < 0$], crosses the real axis at $E^* \approx 10.35$ MeV ($V_0 = -262$ MeV$\cdot$fm$^3$)
close to $E^*=10.991$ MeV, where the $[{^{13}}$N($5/2^+_2) \otimes {\rm p}({\ell_j})]^{J^{\pi}}$ inelastic channel opens. 
This ET dynamics explains the nature of the avoided crossing at $E^* \approx 9.7$ MeV (see Fig. \ref{fig_2a}) as due to the proximity of a 
$J^{\pi}=2^+$ EP with significant  $\ell=0$ coupling to the $[{^{13}}$N($5/2^+_2) \otimes {\rm p}({\ell_j})]^{J^{\pi}}$ channel.
The loop seen in the ET dynamics is due to the opening of  higher-energy decay channels. 

The ET depicted by the green line corresponds to the double poles with capturing asymptotics ({\em i.e.} above real axis). This ET evolves smoothly with the excitation energy and remains a close, but irrelevant, spectator to the configuration mixing of the $2^+_3$ and $2^+_4$ eigenvalues. 

\begin{figure}[t!]
	\includegraphics[width=1.00\linewidth]{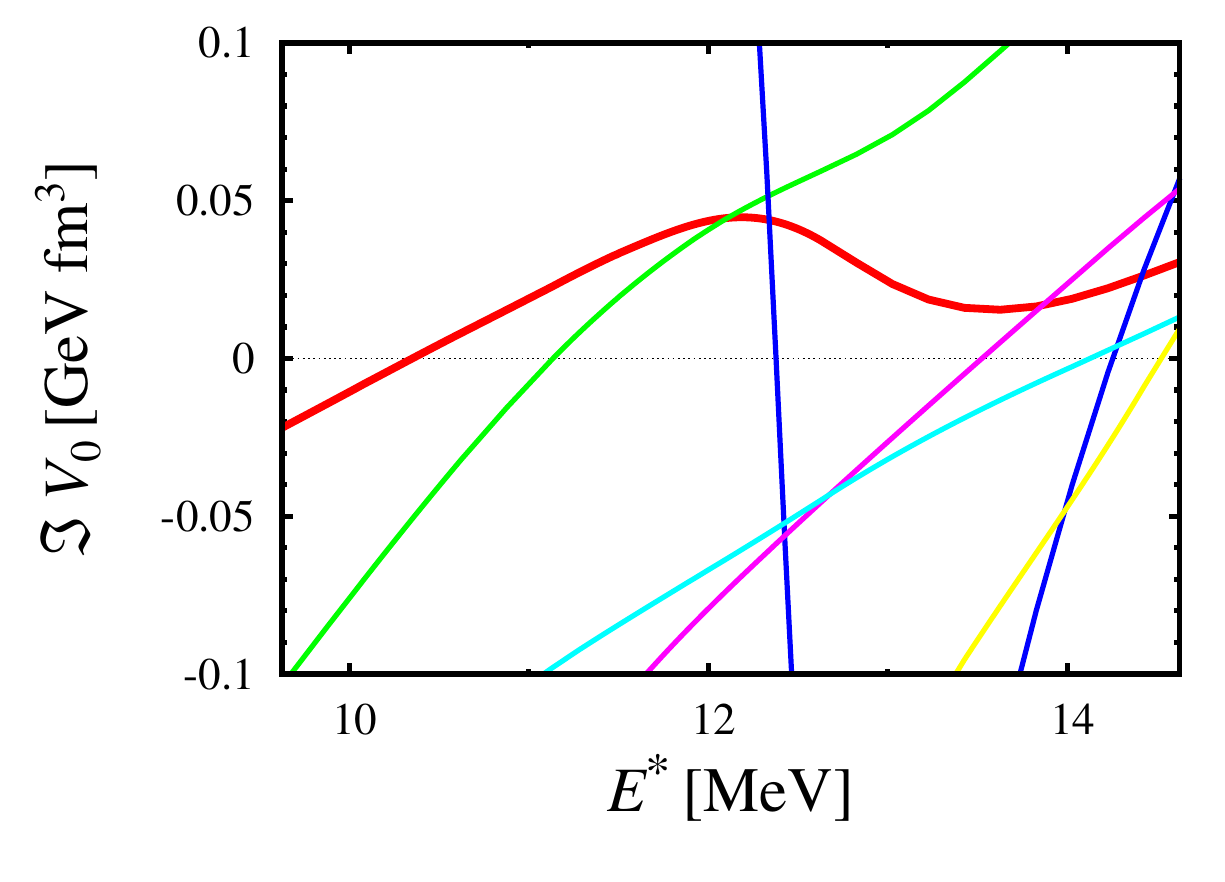}
\caption{ Exceptional threads for $2^+$ eigenstates are displayed in the ${\cal I}m(V_0)$ - $E$ plane. The thick red curve is the exceptional thread responsible for the avoided crossing.}
	\label{fig_2c}
\end{figure}

Another representation of $2^+$ ETs, in the region of small ${\cal I}m(V_0)$, is shown in Fig. \ref{fig_2c}. In the plane of excitation energy $E^*$ and ${\cal I}m(V_0)$ one observes that with increasing excitation energy (and thus higher density of decay channels), more and more ETs cross the physical real axis.  Close to the crossing points, the influence of the double-pole singularity on the configuration mixing of SM eigenvalues increases and hence the movement of SMEC eigenvalues is strongest. In this high excitation-energy region, one peculiar ET (shown with the blue curve) actually crosses the real-$V_0$ axis twice.  While this pattern is quite complex, there is only one ET, depicted with the thick red curve, that appears at the excitation energy sufficiently close to the $2^+_3$ and $2^+_4$ eigenvalues to be relevant for their avoided crossing.

\subsubsection{$1^-$ resonances}

\begin{figure}[t!]
	\includegraphics[width=1.00\linewidth]{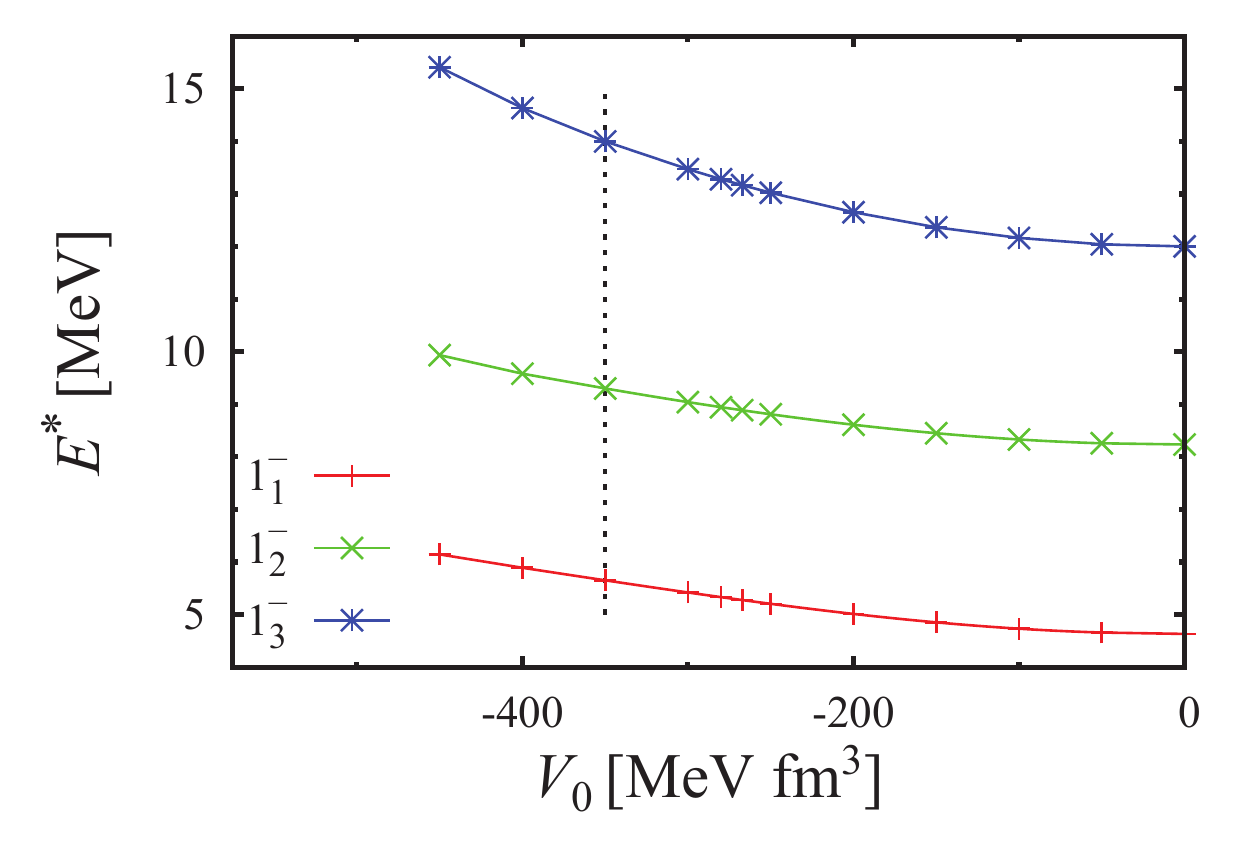}
\caption{SMEC spectrum of $1^-$ eigenstates in $^{14}$O is  shown as a function of the continuum coupling strength $V_0$.  For more details see the caption of Fig. \ref{fig_2a} and the discussion in text.}
	\label{fig_xa}
\end{figure}

Figure \ref{fig_xa} shows the spectrum of three $1^-$ eigenvalues of the SMEC plotted as a function of the (real) continuum coupling constant $V_0$. In this case there no avoided crossings. With increasing  $|V_0|$, the $1^-_1$, $1^-_2$, and $1^-_3$ energies increase gradually. 
The total one-proton decay widths and branching ratios (to different $^{13}$N states) also vary smoothly with $V_0$.

\begin{figure}[t!]
	\includegraphics[width=1.00\linewidth]{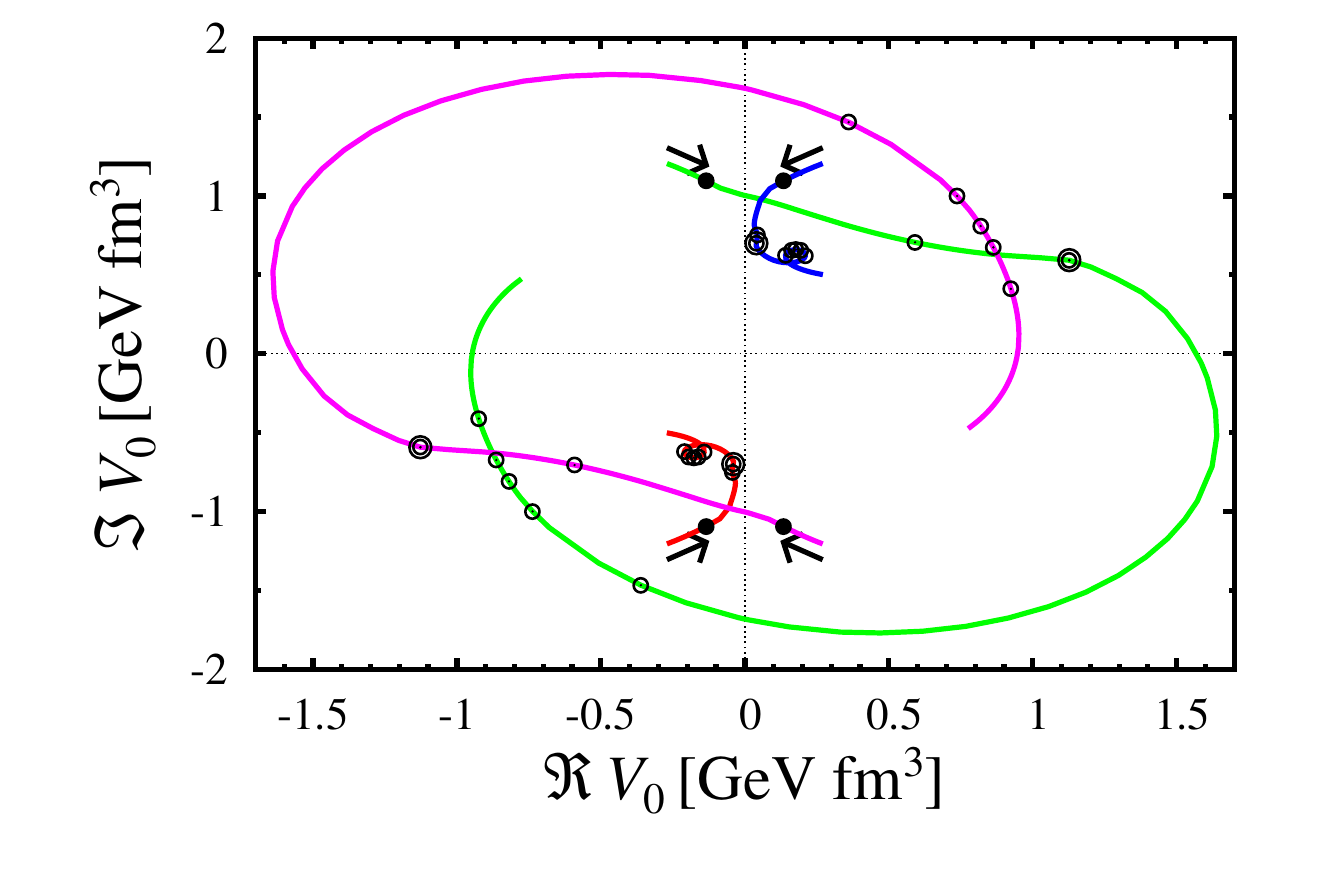}\\
\caption{ Exceptional threads for $1^-$ eigenstates of SMEC in $^{14}$O are shown as a function of the real and imaginary parts of the continuum coupling strength $V_0$. The circles on each curve denote where the excitation energies corresponding to the thresholds for the elastic (solid) and inelastic (open) channels are located. Arrows indicate the direction of increasing excitation energy along each ET.}
	\label{fig_xb}
\end{figure}

This weak configuration mixing among $1^-$ states finds an explanation in the pattern of the most relevant ETs in the ${\cal R}e(V_0)$ - 
${\cal I}m(V_0)$ plane (see Figure \ref{fig_xb}). To follow their evolution, it is necessary to consider both ${\cal R}e(V_0) < 0$ and 
${\cal R}e(V_0) > 0$ half-planes. Below the first decay threshold, the ET corresponding to the poles having a decaying asymptotic is shown as the piece of the short red curve below the filled circle, with the filled circle representing its value at threshold. Its symmetric counterpart for ${\cal R}e(V_0) > 0$, having the capturing asymptotics, is given by the blue curve above its filled circle. These two sets of double poles change rather weakly with increasing  excitation energy, showing a turning near the threshold of the first inelastic channel. These poles are deep inside of the complex plane and hence cannot influence the configuration mixing of $1^-$ eigenstates.

The double-poles which have capturing asymptotics above the first particle threshold are shown by the small piece of the long green curve above its filled circle
at ${\cal R}e(V_0) < 0$.  Its symmetry partner  is given by the small piece of the long magenta curve below its filled circle at ${\cal R}e(V_0) > 0$. Their evolution with  excitation energy is rapid and just above the first decay threshold (filled circles), the ET with the capturing asymptotic (green curve) crosses the 
${\cal R}e(V_0) = 0$ axis and moves into the right half-plane [${\cal R}e(V_0) > 0]$ without changing its asymptotics. Its symmetry partner (magenta curve) makes a move in the opposite direction, entering the half-plane ${\cal R}e(V_0) < 0$. This evolution happens deep in the complex plane for 
$0.5<|{\cal I}m(V_0)|<1.1$ GeV$\cdot$fm$^3$. Above the channel doublet: $[{^{13}}$N($3/2^{-}_1) \otimes {\rm p}(\ell_j)]^{J^{\pi}}$, 
$[{^{13}}$N($5/2^{+}_1) \otimes {\rm p}(\ell_j)]^{J^{\pi}}$, both ETs make a rapid turn passing through the real axis [${\cal I}m(V_0) = 0$] and change their asymptotics. These crossings appear well above a physical range of the continuum coupling constants (${\cal R}e(V_0) \approx 1.5$ GeV$\cdot$fm$^3$) and thus have no influence on the spectra. After opening of all decay channels both of these ETs (depicted by the green and magenta lines) return to their original quadrant of the ${\cal R}e(V_0)$ - ${\cal I}m(V_0)$ plane.

\subsubsection{$0^+$ resonances}

\begin{figure}[t!]
	\includegraphics[width=1.00\linewidth]{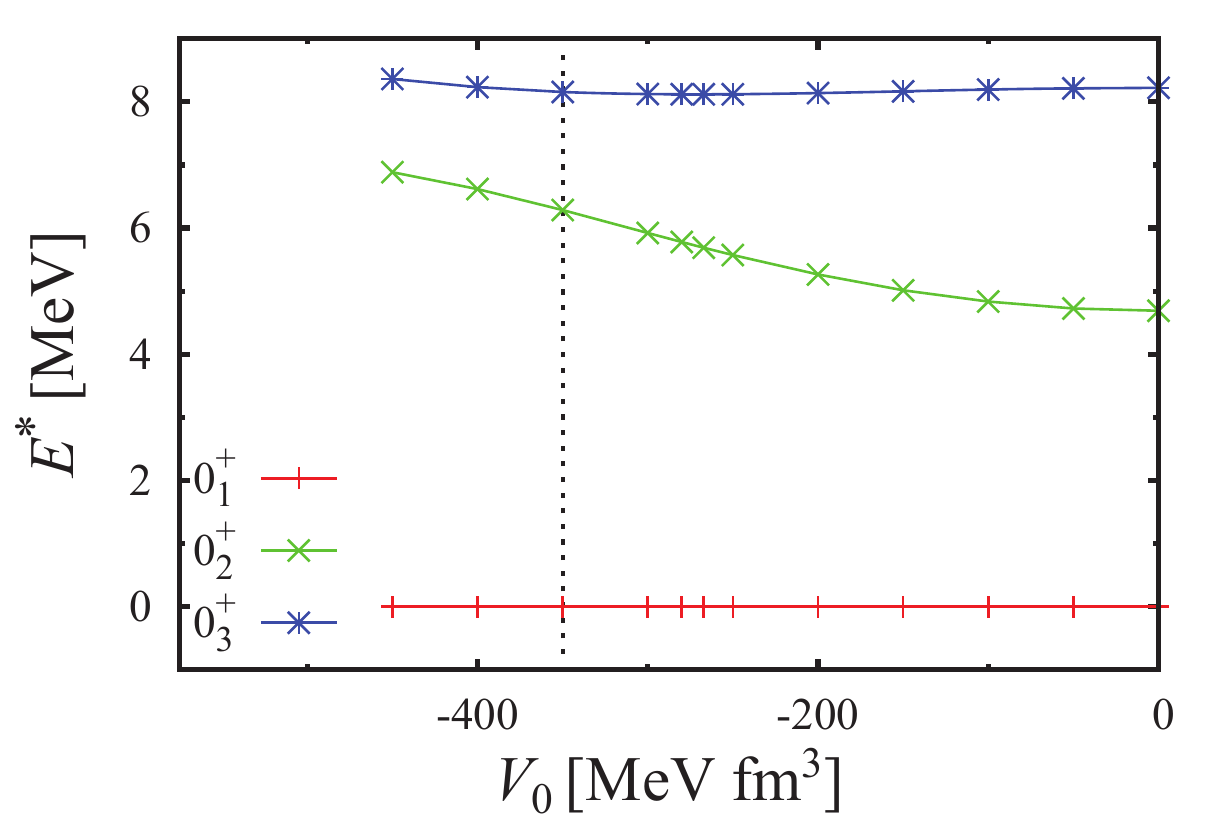}
\caption{The SMEC spectrum of $0^+$ eigenstates in $^{14}$O is shown as a function of the continuum coupling strength $V_0$. For more details see the caption of Fig. \ref{fig_2a} and the discussion in text.}
	\label{fig_4a}
\end{figure}

Figure \ref{fig_4a} shows the spectrum of three $0^+$ eigenvalues as a function of the $V_0$.  The consequence of the fixed-point equations being solved for the ground state (for each $V_0$) is that the eigenenergy of the $0^+_1$ state does not change with $V_0$.
We find that the $0^+_3$ stays approximately constant through the relevant  $V_0$ range.  This means that correlation energy that results from the coupling to the continuum is the same for $0^+_1$ (the ground state) and $0^+_3$. The same cannot be said of $0^+_2$. This state drifts up with increasing  $|V_0|$, relative to the others, as the (negative) correlation energy is less. For very large  $|V_0|$, an avoiding crossing (and thus mixing) of the $0^+_2$ and $0^+_3$ eigenvalues is approached. This crossing (at $V_0 \approx -450$ MeV$\cdot$fm$^3$) corresponds to $E^* \approx 7$ MeV. 

\begin{figure}[t!]
	\includegraphics[width=1.00\linewidth]{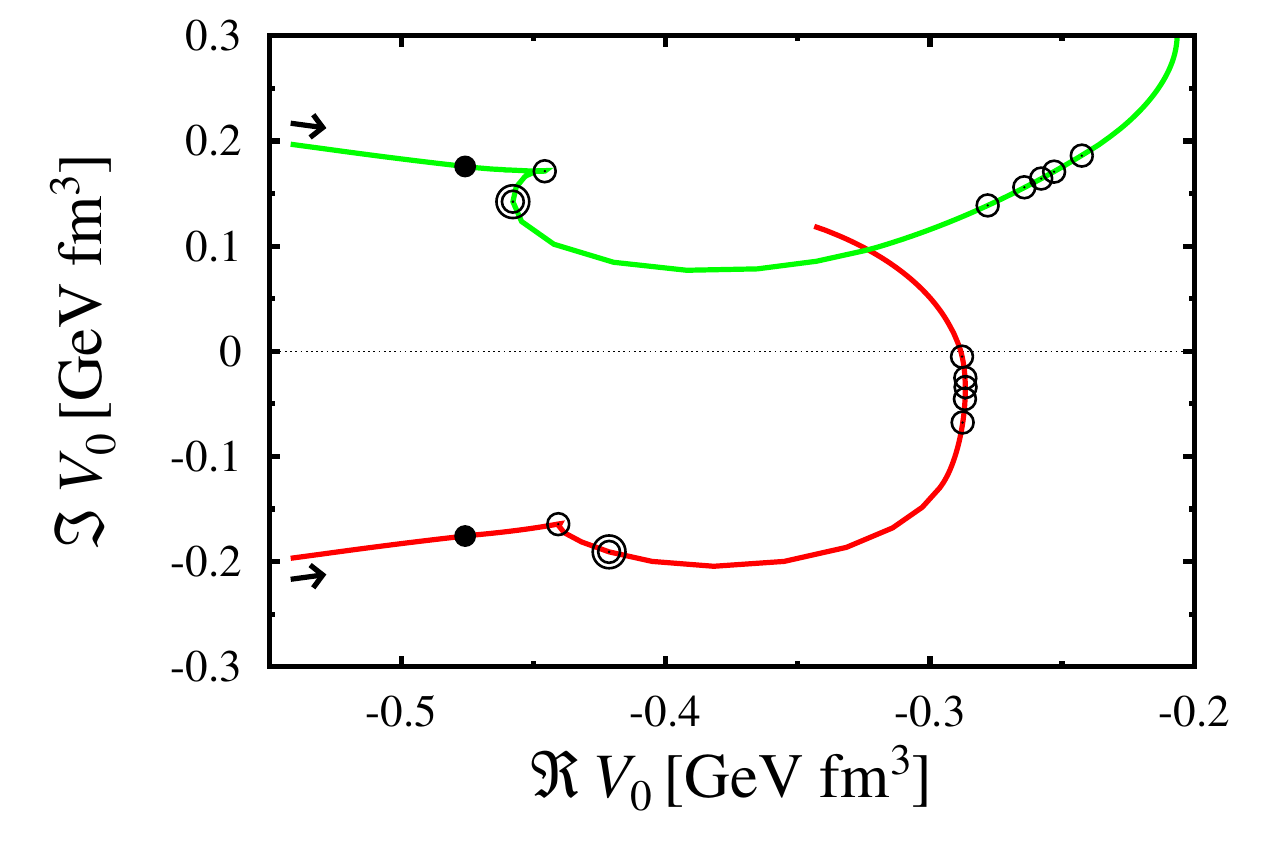}
\caption{ Exceptional threads for $0^+$ eigenstates of the SMEC in $^{14}$O are shown as a function of the real and imaginary parts of the continuum coupling strength $V_0$. The circles on each curve denote where the excitation energies corresponding to the thresholds for the elastic (solid) and inelastic (open) channels are located. Arrows indicate the direction of increasing excitation energy along each ET.}
	\label{fig_4b}
\end{figure}

Figure \ref{fig_4b} displays the two relevant ETs for this avoided crossing. In this case the ETs are both close to the real axis and
have both decaying and capturing asymptotics.
The dynamics of these ETs is caused by the  $\ell=0$ coupling to the decay channel 
$[{^{13}}$N($1/2^{+}_1) \otimes {\rm p}(\ell_j)]^{J^{\pi}}$ which opens at 6.991 MeV.

The ET shown in red, crosses the real axis at $V_0 \approx -288.4$ MeV$\cdot$fm$^3$ as a result of opening of higher lying channels:  $[{^{13}}$N($K^{\pi}) \otimes {\rm p}(\ell_j)]^{J^{\pi}}$ with $K^{\pi}=$$5/2^+_2$, $3/2^+_1$, $7/2^+_1$, $5/2^-_1$, and $3/2^+_2$.
The excitation energy associated with the value of $V_0$ when this ET passes through the physical region near the real axis is very high ($E^* = 12.65$ MeV), well above where these states actually exist.  Thus, as in the case of the  $1^-$ states, strong  mixing due to the proton continuum is not expected for the  $0^+$ states. Nevertheless,  in the relevant range of the continuum coupling strength, the $0^+_1$ and $0^+_3$ resonances drift down fastest relative to other resonances. In particular, the eigenstate $0^+_3$ gains $\approx 1.5$ MeV in correlation energy with respect to $0^+_2$ state at $V_0 = -350$ MeV$\cdot$fm$^3$. 


\section{Discussion} 

The role of the continuum is important for the $^{14}$O states investigated. Specifically,  the exceptional points strongly influence the spectrum and structure of low-energy resonances. Fortunately the continuum coupling acts differently on each of the $\{J^{\pi}\}$ resonance sets and, within a given set, acts selectively on certain states depending on the location of branch points ({\em i.e.} decay thresholds) and double poles of the scattering matrix. The latter do not vary in a systematic way from one nucleus to another. From one point of view this poses a tremendous challenge for the microscopic nuclear theory vis-a-vis the microscopic determination of effective nucleon-nucleon interaction.  From another point of view, with data that are sufficiently discriminatory, the continuum coupling constant can be fixed (for a given nucleus). The case studied here $^{14}$O seems to just such a case.

Figure~\ref{fig:level} presents the experimental and the SMEC results side by side. A search through $V_0$ for a doublet partner for $2^+_2$, finds that only the third $0^+$ state is a serious candidate. One of the two states in the doublet must have a branching ratio to the 1/2$^+$ $^{13}$N excited state of greater than 2.5\% to produce the observed yield in the 2$p$+$^{12}$C channel. 

At the implied coupling strength (which gives $2^+_2$ and $0^+_3$ degeneracy), the $2^+_3$ level coincides with the observed peak at 9.755(10)~MeV. Its  branching ratio  to the negative-parity $^{13}$N ground state is at most 17\%. The SMEC prediction for this branch is 1\%. Most of the remaining decay strength is to the 3/2$^-$, 5/2$^+$ doublet, with decays to the 1/2$^+$ level accounting for only 6.3(9)\%. The SMEC prediction for this branch is $\approx 20\%$.  The 2$^+\rightarrow$ 5/2$^+$ decay path produces a $\cos^4(\theta_{pp})$ term 
in the $\theta_{pp}$ distribution as observed in the data.

The observed structure at 8.787(13) MeV must be $1^-$ as the other unmatched resonances in this energy region: $2^-_2$ and $3^+_1$, exhibit decay patterns which disagree with the experimental findings. Several spin-parity possibilities exist for the highest-excitation-energy peak observed in the present work [11.195(30)~MeV] with the best candidate being the 2$^+_4$ level, but as it is above the alpha threshold (for which the continuum coupling is not considered in this work) there is some uncertainty.

This work does not undertake a study of the mirror of $^{14}$O, {\em i.e.} $^{14}$C. The nucleon and alpha thresholds are shifted up (in going to $^{14}$C) by $\approx 3.54$ MeV and $\approx 1.89$ MeV, making the continuum coupling problem wholly different. However the present work does highlight one very interesting issue in the mirror comparison. The position of  $0^+_2$ is $\approx 0.68$ MeV higher in $^{14}$C than in $^{14}$O while  $0^+_3$ (with our assignment in $^{14}$O) is shifted up $\approx 2.0$ MeV. Not having done the calculations for $^{14}$C, and not having the capability to include coupling to the $\alpha$ channel, we cannot provide an answer for these significant and largely different shifts. However this subject cannot be dropped without making some comments.

There are two mechanisms contributing to these relative shifts. The first mechanism is the well known Thomas-Ehrmann effect which is related to the difference of Coulomb energies in mirror states and hence, is of a geometrical nature. Its relevance is primarily limited to states with significant $s$-orbit contribution as these can expand when a relevant decay threshold is passed in one of the mirror partners. The second mechanism results from the coupling of SM states to decay channels. This continuum coupling may provide both strong energy shifts and collective modifications of the involved wave functions depending on the nature of the matrix elements and a distance from the decay threshold.  In our study of $^{14}$O, the continuum coupling prevails for the $0^+$ states.  Indeed   the coupling to continuum  in $^{14}$O shifts down $0^+_3$ relative to $0^+_2$ by $\approx 1.5$ MeV at $V_0 = -350$ MeV$\cdot$fm$^3$.
On the other hand, the contribution of the $s^2$ configurations are $<1\%$, $\approx 59\%$, $\approx 10\%$ for the $0^+_1$, $0^+_2$, and $0^+_3$ states, respectively. Therefore without an additional source of mixing, the strong mirror shift between the $0^+_3$ states of $^{14}$C and $^{14}$O cannot be induced by the standard Thomas-Ehrmann effect.

\section{CONCLUSIONS}

The A = 14 isobaric chain contains many fascinating issues. One of these is the perturbation of the $^{14}$O shell-model spectrum by open decay channels. Only the ground state is particle bound and proton decay to the first four states of $^{13}$N are the only open channels until the $\alpha + ^{10}$C channel opens at 10.12 MeV.
This work is a first attempt to study this nuclide, in this excitation widow, with insight gained from invariant-mass measurements combined with the SMEC, a continuum-cognizant shell model. 
Experimentally, $^{14}$O excited states are produced from knockout reactions with a fast $^{15}$O beam. The decay produced from 1$p$ and 2$p$ decay are detected with the HiRA array and used to construct invariant mass-distributions. From the momentum correlations observed in the 2$p$ decays, sequential decay pathways through a number of possible $^{13}$N intermediate states can be inferred and branching ratios extracted. 
The model study, in and of itself, indicated that exceptional points lurk in the Hilbert space, making this type of project challenging. 

This experimental information on energies, widths and branching ratios when compared to the same information from the calculations, provides discrimination between the possible spin and parities. We found that the previously known 2$^+_2$ state forms a doublet with a previously unknown state. If the strength of the coupling of shell-model states to the continuum is varied, one finds that the only possible partner is the 0$^+_3$ level.  Also the assignment of 2$^+_3$ becomes clear leaving the only observed state between the second and third 2$^+$ states to have a 1$^-$ assignment. Thus all observed levels between the proton and alpha thresholds are assigned. 

The decay of the 2$^+_3$ excited state demonstrates an interesting example of interference in sequential 2$p$ decay. The sequential decay paths for this state are largely through a pair of degenerate states in $^{13}$N of opposite parity (3/2$^-$ and 5/2$^{+}$). The interference between these two decays paths results in a $p$-$p$ relative angle ($\theta_{pp}$) distribution which is no longer symmetric about $\theta_{pp}$=90$^\circ$, the expectation for most sequential 2$p$ decays.  

Finally, our assignment of the $0^+_3$ state in $^{14}$O motivates a detailed comparison between $^{14}$O and its mirror $^{14}$C with models that can include coupling to all relevant open channels. 

\begin{acknowledgments}
This work was supported by the U.S. Department of Energy, Division of Nuclear Physics under grant No. DE-FG02-87ER-40316, No. DE-FG02-04ER-41320, and No. DE-SC0014552 and by the National Science Foundation under Grant No. PHY-0606007 
and by the COPIN and COPIGAL French-Polish scientific exchange programs.  
K.W.B. was supported by a National Science Foundation Graduate Fellowship under Grant No. DGE-1143954 and J.M. was supported by a Department
of Energy National Nuclear Security Administration Stewardship Science Graduate Fellowship under cooperative Agreement Number DE-NA0002135.
\end{acknowledgments}

   
%

   \end{document}